\documentclass[10pt,journal,compsoc]{IEEEtran}
\ifCLASSOPTIONcompsoc
  \usepackage[nocompress]{cite}
\else
  \usepackage{cite}
\fi
\ifCLASSINFOpdf
\else
\fi

\usepackage{amsmath,amsfonts}
\usepackage{array}
\usepackage[caption=false,font=normalsize,labelfont=sf,textfont=sf]{subfig}
\usepackage{textcomp}
\usepackage{stfloats}
\usepackage{url}
\usepackage{verbatim}
\usepackage{graphicx}
\usepackage{cite}

\usepackage{algorithm}  
\usepackage{algorithmicx}  
\usepackage{algpseudocode}

\usepackage{color}

\usepackage[skip=0.333\baselineskip]{caption}
\usepackage{booktabs,tabularx,ragged2e}
\usepackage{multirow}
\usepackage{graphicx}
\usepackage[table]{xcolor}

\newcolumntype{Y}{>{\hsize=.7\hsize\RaggedRight\arraybackslash}X}
\newcolumntype{a}{>{\columncolor{lightgray}}Y}
\newcolumntype{C}[1]{>{\columncolor{lightgray}}p{#1}} 

\newtheorem{rem}{Remark}

\usepackage{amsmath,amsfonts,bm}

\def\rvd{{\mathbf{d}}}

\def\rvu{{\mathbf{i}}}

\def\rvn{{\mathbf{n}}}

\def\rvu{{\mathbf{u}}}

\def\rvx{{\mathbf{x}}}
\def\rvy{{\mathbf{y}}}
\def\rvz{{\mathbf{z}}}

\def\rmI{{\mathbf{I}}}

\DeclareMathAlphabet{\mathsfit}{\encodingdefault}{\sfdefault}{m}{sl}
\SetMathAlphabet{\mathsfit}{bold}{\encodingdefault}{\sfdefault}{bx}{n}

\def\gB{{\mathcal{B}}}

\def\gD{{\mathcal{D}}}

\def\gN{{\mathcal{N}}}

\def\gT{{\mathcal{T}}}
\def\gU{{\mathcal{U}}}

\newcommand{\E}{\mathbb{E}}

\newcommand{\KL}{D_{\mathrm{KL}}}

\begin{document}
%
\title{Learn from Unpaired Data for Image Restoration: A Variational Bayes Approach}
%
%
%
%

\author{Dihan Zheng, Xiaowen Zhang, Kaisheng Ma, Chenglong Bao$^{*}$ 
\IEEEcompsocitemizethanks{
	
	\IEEEcompsocthanksitem Dihan Zheng is with the Yau Mathematical Sciences Center, Tsinghua University, Beijing 100084, China \protect\\
	(e-mail: zhengdh19@mails.tsinghua.edu.cn)
	
	\IEEEcompsocthanksitem Xiaowen Zhang is with the Hisilicon, Shanghai 300060, China \protect\\
	(e-mail: zhangxiaowen9@hisilicon.com)
	
	\IEEEcompsocthanksitem Kaisheng Ma is with the Institute for Interdisciplinary Information Science, Tsinghua University, Beijing 100084, China \protect\\
	(e-mail: kaisheng@mail.tsinghua.edu.cn)
	
	\IEEEcompsocthanksitem Chenglong Bao is with the Yau Mathematical Sciences Center, Tsinghua University, Beijing 100084, China, and Yanqi Lake Beijing Institute of Mathematical Sciences and Applications, Beijing 101408, China \protect\\
	(e-mail: clbao@mail.tsinghua.edu.cn)
	
	\IEEEcompsocthanksitem $^*$ Corresponding author.
	}
\thanks{Manuscript received; revised.}}

\IEEEtitleabstractindextext{%
\begin{abstract}
	Collecting paired training data is difficult in practice, but the unpaired samples broadly exist. Current approaches aim at generating synthesized training data from unpaired samples by exploring the relationship between the corrupted and clean data. This work proposes LUD-VAE, a deep generative method to learn the joint probability density function from data sampled from marginal distributions. Our approach is based on a carefully designed probabilistic graphical model in which the clean and corrupted data domains are conditionally independent. Using variational inference, we maximize the evidence lower bound (ELBO) to estimate the joint probability density function. Furthermore, we show that the ELBO is computable without paired samples under the inference invariant assumption. This property provides the mathematical rationale of our approach in the unpaired setting. Finally, we apply our method to real-world image denoising, super-resolution, and low-light image enhancement tasks and train the models using the synthetic data generated by the LUD-VAE. Experimental results validate the advantages of our method over other approaches.
\end{abstract}

}

\maketitle

\IEEEdisplaynontitleabstractindextext

%
\IEEEpeerreviewmaketitle

\begin{figure*}
	\centering
	\begin{tabular}{c@{\hspace{0.01\linewidth}}c@{\hspace{0.01\linewidth}}c@{\hspace{0.01\linewidth}}c@{\hspace{0.01\linewidth}}c}
		\includegraphics[width=0.183\linewidth]{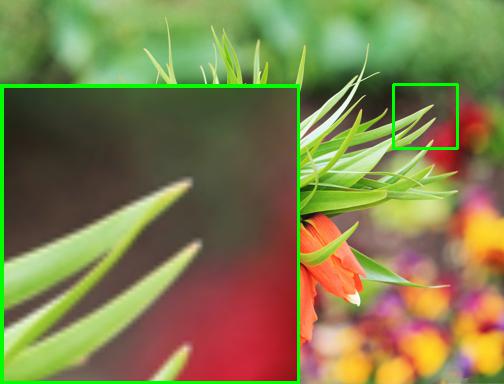}   &
		\includegraphics[width=0.183\linewidth]{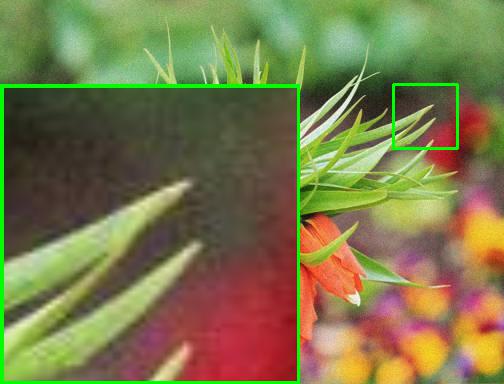}   &
		\includegraphics[width=0.183\linewidth]{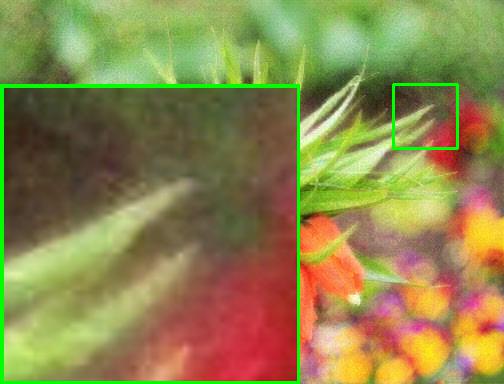} &
		\includegraphics[width=0.183\linewidth]{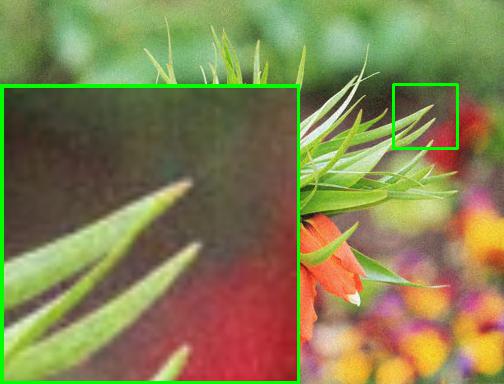} &
		\includegraphics[width=0.183\linewidth]{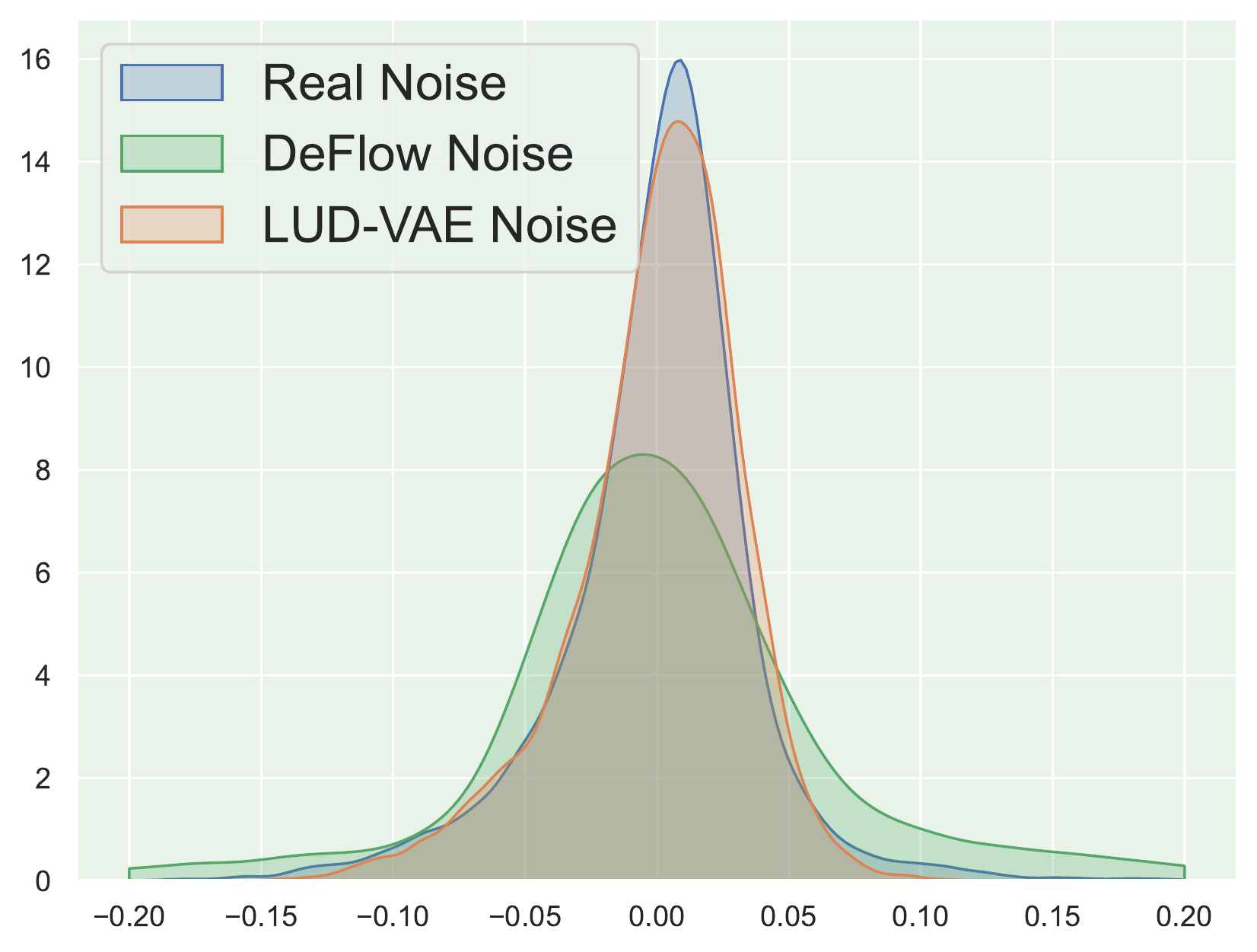} \\
		
		\includegraphics[width=0.183\linewidth]{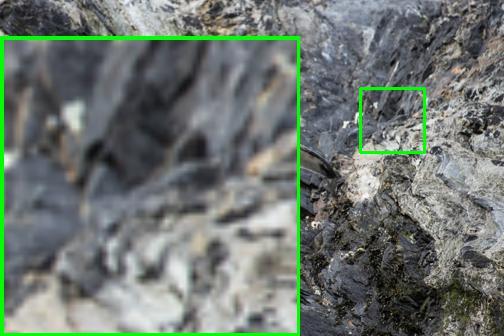}   &
		\includegraphics[width=0.183\linewidth]{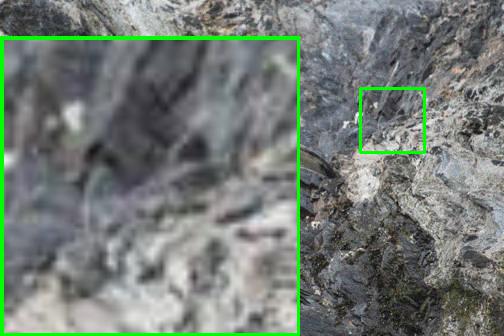}   &
		\includegraphics[width=0.183\linewidth]{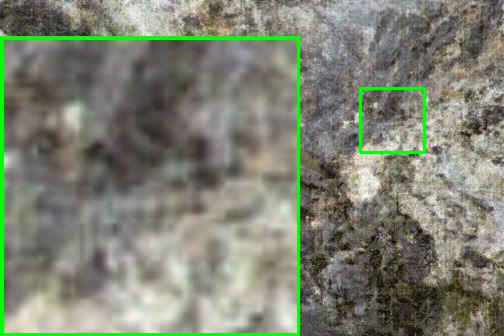} &
		\includegraphics[width=0.183\linewidth]{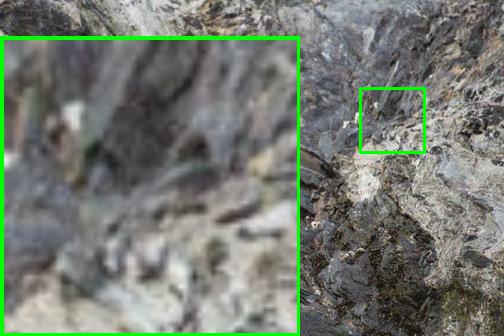} &
		\includegraphics[width=0.183\linewidth]{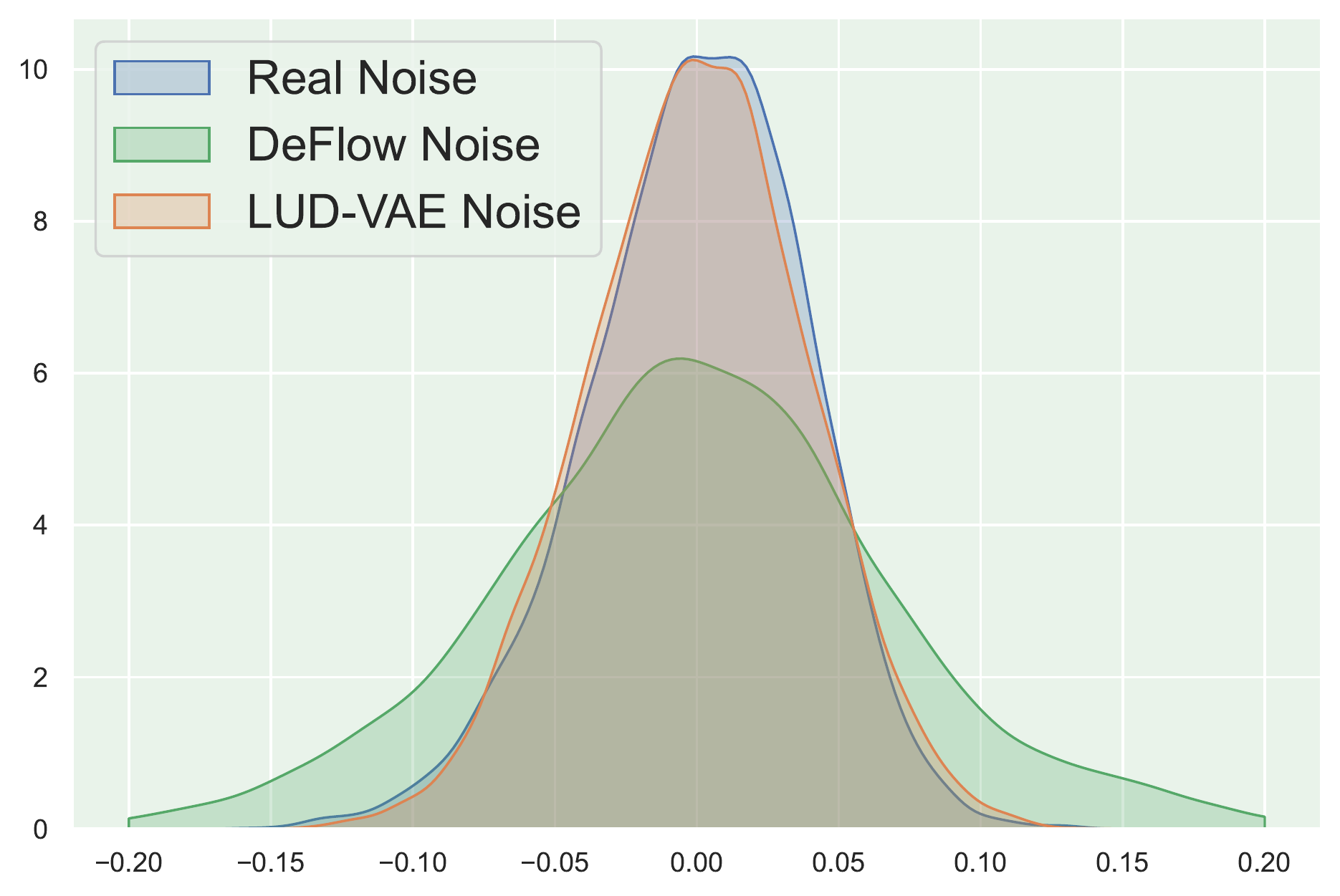} \\
		
		\includegraphics[width=0.183\linewidth]{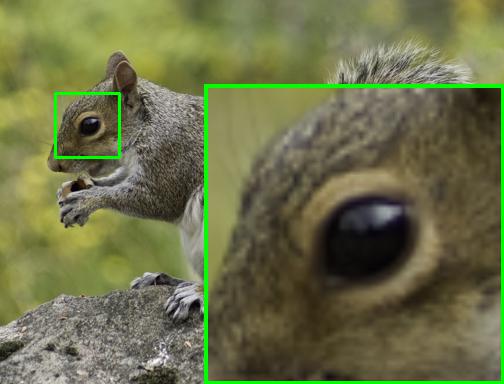}   &
		\includegraphics[width=0.183\linewidth]{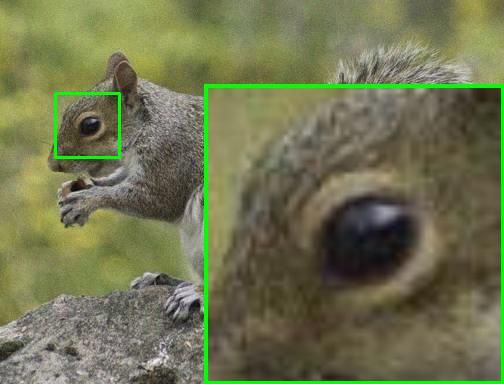}   &
		\includegraphics[width=0.183\linewidth]{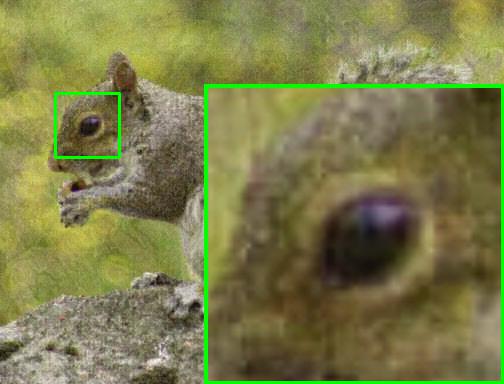} &
		\includegraphics[width=0.183\linewidth]{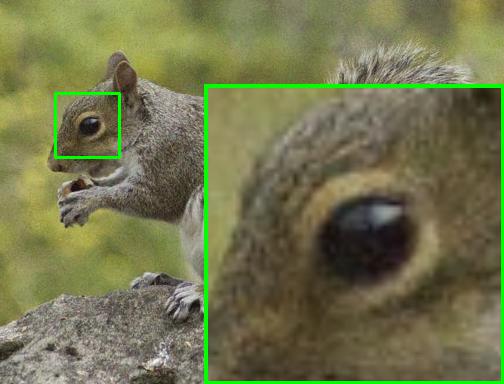} &
		\includegraphics[width=0.183\linewidth]{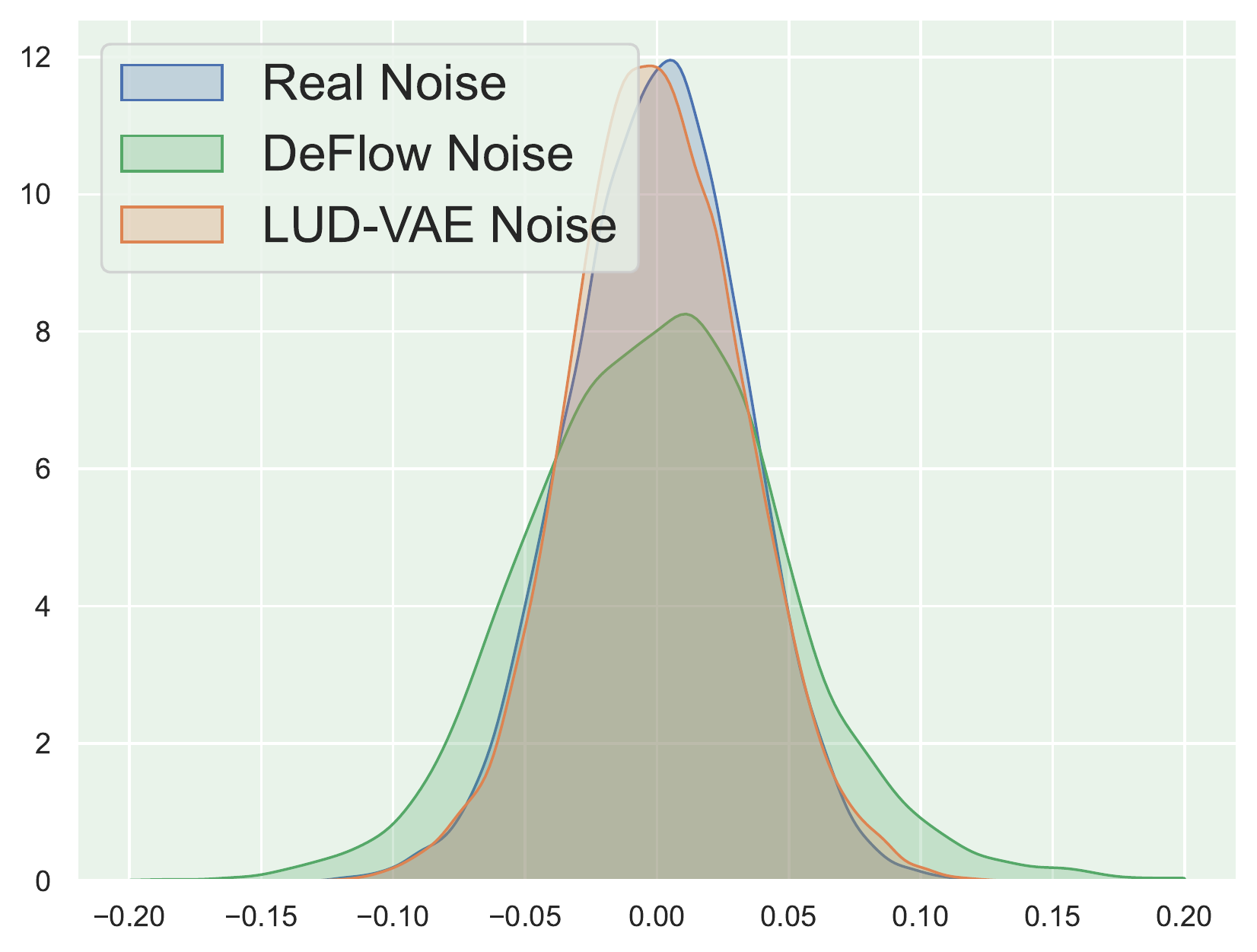} \\
		\small{Clean image} & \small{Real noise} & \small{Noise from DeFlow~\cite{wolf2021deflow}} & \small{Noise from LUD-VAE} & \small{Noise distribution}\\
	\end{tabular}
	\caption{Synthetic noisy images obtained by DeFlow and LUD-VAE learned from unpaired noisy-clean dataset AIM2019~\cite{lugmayr2019aim}. Patches of $32 \times 32$ from each images are chosen for comparison, the noise distribution in these patches are shown.}
	\label{intro}
\end{figure*}

\IEEEraisesectionheading{\section{Introduction}}
Image restoration aims to recover the underlying clean image $\rvx$ from the corrupted observation $\rvy$,
\begin{equation}
	\rvy = \gT(\rvx) + \rvn,
\end{equation}
where $\rvn$ represents the noise, and $\gT$ represents the degradation operation. This task is one of the fundamental problems in computer vision and has been extensively studied for decades~\cite{buades2005non,dabov2007image,gu2014weighted}. In recent years, deep learning has achieved astonishing success in image restoration problems, such as image denoising~\cite{zhang2017beyond,zhang2018ffdnet,guo2019toward} and super-resolution~\cite{dong2014learning,dong2016accelerating,wang2018esrgan,lugmayr2020srflow}. However, the success of these methods requires large quantities of paired training data, and the restoration performance is sensitive to the degradation types~\cite{abdelhamed2018high,castillo2021generalized,mohan2019robust}. For example, one Gaussian denoising network usually performs poorly for real-world noisy images due to the noise discrepancy between Gaussian noise and real-world noise~\cite{zheng2020unsupervised}. Meanwhile, collecting paired training data for real-world image restoration is cumbersome and expensive due to the complex camera image signal processing (ISP) pipeline~\cite{abdelhamed2019noise,gow2007comprehensive,hasinoff2010noise}. The aforementioned problems make real-world image restoration a challenging task. On the other hand, unpaired data broadly exists and is easily accessible in many situations. For example, it is easy to obtain many images of different resolutions or noisy and clean images through the internet~\cite{chen2018image}. Consequently, designing deep learning methods with unpaired data is of significant research importance and deserves deep exploration.

Currently, two common strategies are along this line. One is the unsupervised image restoration methods with a single corrupted image~\cite{ulyanov2018deep,quan2020self2self,zheng2020unsupervised} or corrupted image dataset~\cite{wu2020unpaired,laine2019high}, which do not take clean images into consideration. As a result, those methods are either time-consuming or inferior to supervised methods. The other strategy is to learn the degradation model from the unpaired datasets. After learning a generative model to construct the synthetic paired training data, it trains an image restoration model using conventional supervised deep learning methods. The main difficulty of these methods is to develop effective methods so that the synthetic paired data is close to the underlying paired data. Current methods mainly adopt generative adversarial networks (GANs)~\cite{goodfellow2014generative} and draw on the idea from cycle-consistency constraint in Cycle-GAN~\cite{bulat2018learn,lugmayr2019unsupervised,zhu2017unpaired} and domain adversarial training~\cite{bell2019blind,fritsche2019frequency,wei2021unsupervised}. However, these methods often require careful adjustment of different losses, and the heuristic constraint of cycle consistency is too weak for this problem and lacks theoretical rigorousness~\cite{wolf2021deflow,chu2017cyclegan}. More importantly, those GAN-based methods usually obtain a deterministic mapping while ignoring the randomness in the degradation generation process. Recently, the DeFlow~\cite{wolf2021deflow} method, which models the unpaired degradation process using a conditional flow model, has shown promising performance in super-resolution. In Figure~\ref{intro}, it shows that the noise distribution generated by Deflow does not well match the true noise distribution. 

In this work, we propose LUD-VAE\footnote{LUD-VAE stands for Learning from Unpaired Data using Variational Auto-Encoder.}, a variational auto-encoder (VAE) based degradation modeling method with unpaired training data. Mathematically, given $\rvx\sim p(\rvx)$ and $\rvy\sim p(\rvy)$, our goal is to approximate the joint distribution $p(\rvx,\rvy)$. In general, as shown in~\cite{wolf2021deflow}, this task is difficult and has no unique solution. Thus, it motivates us to carefully design a computable generative graph that well approximates a reasonable solution. More specifically, this graph consists of two independent latent variables: $\rvz$, which encodes the image information, and $\rvz_\rvn$, which encodes the degradation information. The generation relationship is designed as follows: $\rvx$ is generated from $\rvz$; $\rvy$ is generated from $\rvz$ and $\rvz_\rvn$. Using the idea from VAE, we introduce an encoder network for the inference and a decoder network for the generative process so that we can model the conditional distribution $p(\rvy | \rvx)$. In addition, we impose the inference invariant condition on $\rvz$ that requires the same latent representations of paired $\rvx$ and $\rvy$ and show that the ELBO can be computed via unpaired corrupted and clean data. This property gives us a transparent and explainable loss for training the networks. The Figure~\ref{intro} demonstrates that the proposed method can learn the noise distribution accurately. Finally, we apply the LUD-VAE model to the problem of real-world image denoising, super-resolution, and low-light image enhancement. LUD-VAE is used to learn the degradation model with unpaired data and synthesize paired training data for the downstream supervised models. We test the performance of LUD-VAE on two real-world super-resolution datasets: AIM2019~\cite{lugmayr2019aim} and NTIRE2020~\cite{lugmayr2020ntire}, two real-world image denoising dataset: SIDD~\cite{abdelhamed2018high} and DND~\cite{plotz2017benchmarking}, and one low-light image enhancement dataset: LOL~\cite{wei2018deep}. Experimental results show that the proposed LUD-VAE model outperforms GAN-based approaches. Compared with the recent flow-based method, our model achieves comparable results with much fewer parameters. Our main contributions are summarized as follows.

\begin{itemize}
	\item We propose a degradation modeling method, called the LUD-VAE, by constructing a new generative graph. Using the VAE framework, we derive an explainable loss function that decouples the dependency between noisy and clean images. The work is the first attempt to model the underlying paired data distribution for image restoration from the VAE perspective to the best of our knowledge.
	\item We design a hierarchical structure for LUD-VAE to effectively learn the degradation process with an economic network. Moreover, the model is self-supervised and does not require pre-trained networks.
	\item Experimental results in the task of real-world image denoising, super-resolution, and low-light image enhancement validate the advantages of the proposed LUD-VAE.
\end{itemize}

\section{Related work}
{\noindent \bf Deep image restoration.} In recent years, deep learning based image restoration methods have significantly improved performance over traditional methods. In additive white Gaussian noise (AWGN) removal task, DnCNN~\cite{zhang2017beyond} uses deep convolutional networks to predict noise from the noisy images. Further, FFDNet~\cite{zhang2018ffdnet} takes the noise level as network input to deal with images with different noise levels. Recently, researchers have designed different structures to improve denoising performance~\cite{chen2018deep,jia2019focnet,zhang2021plug}. While these methods have achieved state-of-the-art performance on the synthetic datasets, they still require numerous paired training data to learn the restoration process. In real-world image restoration tasks, collecting paired data is difficult due to the sophisticated imaging process. Some real-world image restoration datasets such as image denoising~\cite{plotz2017benchmarking,abdelhamed2018high,anaya2018renoir,abdelhamed2019ntire,guo2019toward} and super-resolution~\cite{lugmayr2019unsupervised,cai2019ntire,cai2019toward}. They use statistical methods to synthesize clean images from multiple noisy observations, requiring careful setups and procedures. With these datasets, many methods are proposed for real-world image restoration, such as using various attention modules~\cite{anwar2019real,kim2019grdn,zhang2018image}, vision transformers~\cite{liang2021swinir,zamir2021restormer,wang2021uformer}, and GANs~\cite{ledig2017photo,wang2018esrgan}. 
There are also some unsupervised or self-supervised methods, which reduce the requirements for training datasets. N2N~\cite{lehtinen2018noise2noise} proposes to use paired noise images to learn an image denoising network. Subsequently, N2V~\cite{krull2019noise2void} and N2S~\cite{batson2019noise2self} further reduce this requirement. They use a "blind spot" network to learn a denoising model directly from the noise data set without clean images. However, these methods need to assume that the image noise is spatially independent, limiting their application to real data. 

{\noindent \bf Unpaired degradation modeling.} Learning the degradation model from unpaired data can be considered the image-to-image transfer task, a long-standing problem in computer vision. Most of the existing works employ the GANs~\cite{goodfellow2014generative}, mainly using cycle-consistency~\cite{bulat2018learn,lugmayr2019unsupervised,ye2021closing,lu2019unsupervised} proposed in Cycle-GAN~\cite{zhu2017unpaired} and domain adversarial~\cite{bell2019blind,fritsche2019frequency,wei2021unsupervised} to characterize the conditional relationship between $p(\rvx)$ and $p(\rvy)$. The core of these methods is to disentangle the degradation and content information, where they choose to design explicit regularization losses, such as domain adversarial loss, cycle consistent loss, perceptual loss, and KL divergence loss. However, these strategies usually lack theoretical guarantees and need elaborate fine-tuning of those losses~\cite{wolf2021deflow}. Meanwhile, GAN-based methods may have some issues, such as unstable training~\cite{arjovsky2017towards} and mode collapse~\cite{arjovsky2017wasserstein}. Instead, Our model is based on variational inference, each term in our loss function can be derived from the evidence lower bound, and there are no adversarial losses as in GANs. Moreover, the disentanglement of the proposed LUD-VAE is based on the construction of the generative graph and the inference invariant condition. Specifically, our method implements the inference invariant condition using a pre-process operator without imposing extra loss functions. There are also handcrafted methods to synthesize degradation images~\cite{ji2020real}, but they lack generalization ability~\cite{wolf2021deflow,castillo2021generalized}. Recently,~\cite{wolf2021deflow} proposed the DeFlow model, a flow-based degradation modeling method without paired data, which has achieved excellent performance on real-world super-resolution tasks.

{\noindent \bf Unpaired learning with VAEs.} Variational auto-encoder (VAE)~\cite{kingma2013auto} based unpaired learning for degradation modeling is currently less developed; here, we investigate some related topics. \cite{zhao2021unpaired} proposed to train an Energy-Based Model (EBM) in the latent space of a trained VAE to realize image-to-image transfer. However, this method uses Markov Chain Monte Carlo (MCMC) algorithms to sample from the latent space, which leads to slow generation speed. Meanwhile, this method has no theoretical constraints to ensure the rationality of the transformation of $\rvx$ to $\rvy$, which is mainly used for the transformation between human faces or animal images. In~\cite{zheng2020unsupervised}, a single image based unsupervised denoising method with VAEs is proposed, but this method requires training a new network for each image. \cite{prakash2020fully} proposed a dataset based unsupervised denoising algorithm for microscopy images with VAEs, which needs to know the noise distribution and requires multiple re-samplings to boost the performance. In addition, it only verifies the effectiveness for microscopy images and text images, whereas the texture and content of natural images are more complicated. In this work, we learn the degradation model with unpaired data and design an explainable and reasonable loss with successful applications in real-world image restoration tasks.

\section{Our methodology}
In this section, we present our method for learning the unknown degradation model using unpaired noisy and clean images. Formally, assume the data $\{\rvx_i\}$ and $\{\rvy_j\}$ are i.i.d. sampled from $p(\rvx)$ and $p(\rvy)$ respectively, and our goal is to generate paired samples from the conditional distribution $p(\rvy | \rvx)$. In the following context, we assume clean images lie in the source domain and noisy images lie in the target domain.

\begin{figure}
	\centering
	\includegraphics[width=0.5\linewidth]{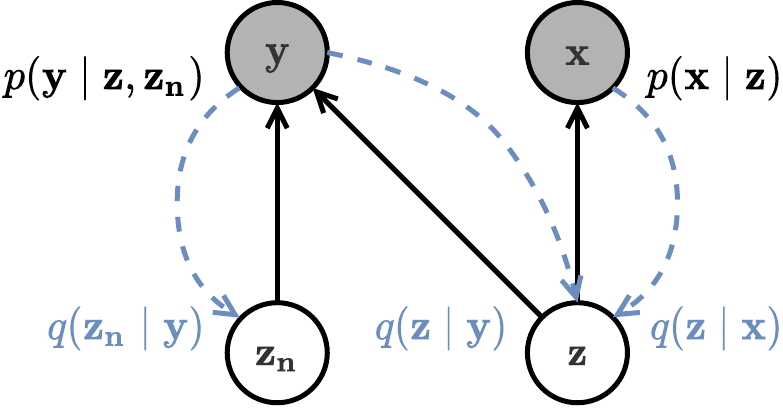}
	\caption{Graphical model of the image generation process. $(\rvx,\rvy)$ are paired clean and noisy images; $(\rvz,\rvz_n)$ are latent variables for generating $(\rvx,\rvy)$.}
	\label{graph_model}
\end{figure}

\begin{figure*}
	\centering
	\includegraphics[width=1.\linewidth]{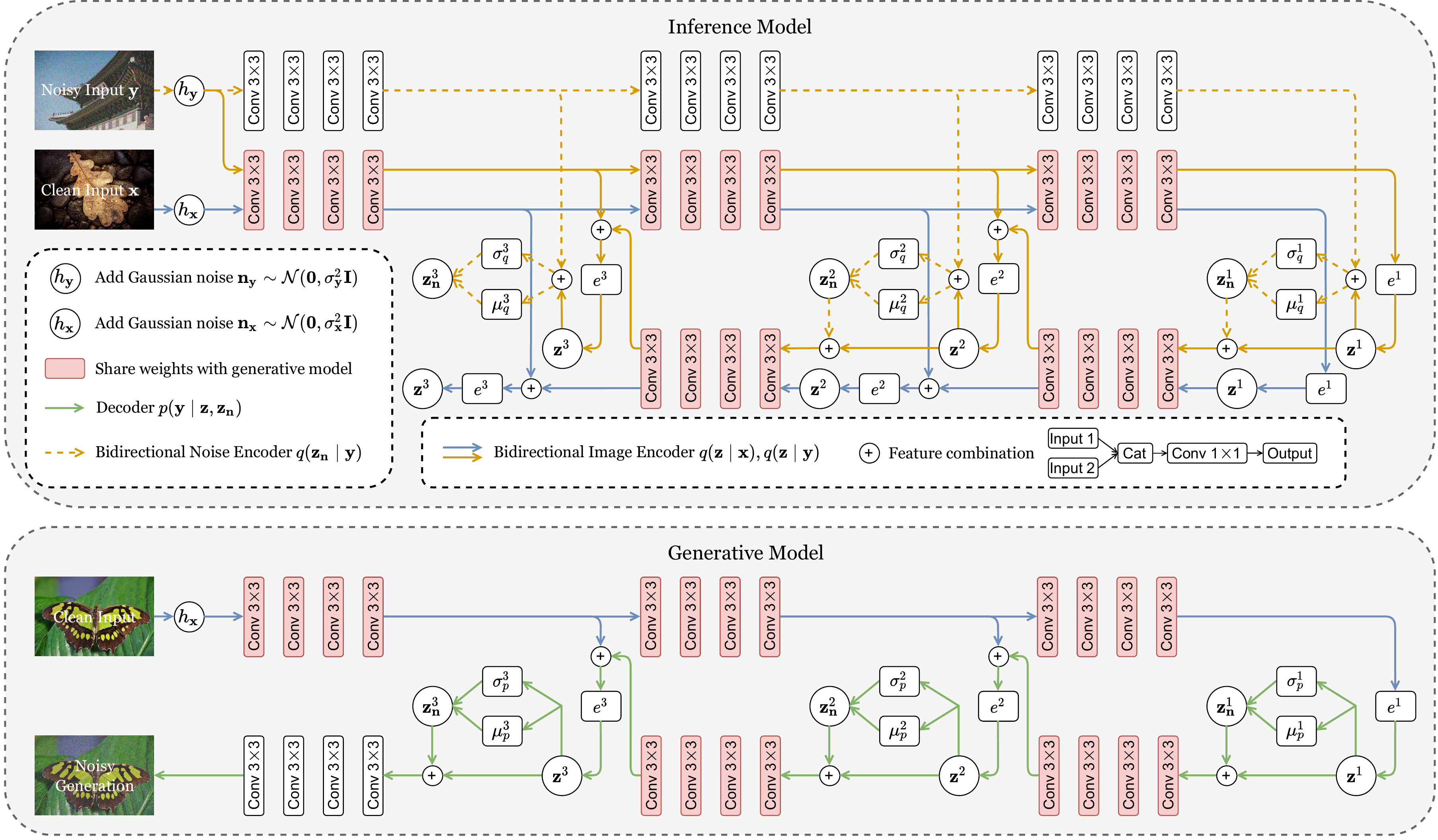}
	\caption{The hierarchical architecture used in LUD-VAE.}
	\label{model_arch_fig}
\end{figure*}

\subsection{Basic idea}
To find the transformation from $\rvx$ to $\rvy$, one straightforward idea is to estimate the conditional density $p(\rvy | \rvx)$. However, it is difficult to model the conditional density function directly due to the lack of paired data. Instead, we consider to model the joint density function $p(\rvx, \rvy)$ in this work. To leverage the information of the unpaired data, our basic idea is to decouple the joint density function into the source domain and target domain. Since the independent assumption for $\rvx,\rvy$ does not hold in practice, we impose the conditional independence by assuming the joint random variable $(\rvx,\rvy)$ has two latent variables: $\rvz$ and $\rvz_\rvn$. For a paired data $(\rvx,\rvy)$ sampled from $p(\rvx, \rvy)$, we assume that the image content and the degradation information are from latent variables $\rvz$ and $\rvz_\rvn$ respectively. See Figure~\ref{graph_model} for the generative graph. Under the above assumptions, the conditional joint distribution becomes
\begin{equation}
	p(\rvx,\rvy | \rvz, \rvz_\rvn) = p(\rvx | \rvz) p(\rvy| \rvz,\rvz_\rvn).
\end{equation}

Assume $\{(\rvx_i, \rvy_i)\}_{i=1}^{N}$ is the paired dataset sampled from $p(\rvx, \rvy | \rvz, \rvz_\rvn)$, where $N$ is the number of samples. Then the conditional log-likelihood is
\begin{equation}
	\begin{aligned}
		&  \sum_{i=1}^{N} \log p(\rvx_i,\rvy_i | \rvz, \rvz_\rvn) \\
		=& \sum_{i=1}^{N} \left( \log p(\rvx_{i} | \rvz) + \log p(\rvy_i | \rvz,\rvz_\rvn) \right) \\
		=& \sum_{i=1}^{N} \log p(\rvx_{i} | \rvz) + \sum_{j=1}^N \log p(\rvy_j | \rvz,\rvz_\rvn),
	\end{aligned}
\end{equation}
where $\{\rvx_i\}_{i=1}^{N}$ and $\{\rvy_j\}_{j=1}^{N}$ are sampled from $p(\rvx|\rvz)$ and $p(\rvy|\rvz,\rvz_\rvn)$ respectively. Thus, it removes the dependence of paired data, and use unpaired $\{\rvx_i\}_{i=1}^{N}$ and $\{\rvy_j\}_{j=1}^{N}$ to evaluate the conditional log-likelihood function. Inspired by the above observation, we can approximate the joint distribution $p(\rvx,\rvy)$ using the VAE framework in the next subsection. 

\subsection{Proposed LUD-VAE method}
To estimate the joint density function $p(\rvx, \rvy)$ with the graphical model given by Figure~\ref{graph_model}, we apply the variational inference framework. 
Note that the log-likelihood function $\log p(\rvx, \rvy)$ has the following decomposition:
\begin{equation}\label{eqn:log-likelihood}
	\begin{aligned}
		\log p(\rvx, \rvy) =& \E_{q(\rvz, \rvz_\rvn| \rvx,\rvy)} \log \frac{p(\rvz, \rvz_\rvn, \rvx,\rvy)}{q(\rvz, \rvz_\rvn| \rvx,\rvy)} \\ 
		&+ \KL\left(q(\rvz, \rvz_\rvn| \rvx,\rvy) \| p(\rvz, \rvz_\rvn| \rvx,\rvy)\right),
	\end{aligned}
\end{equation}
where $\KL$ is the Kullback–Leibler (KL) divergence. Since the second term in \eqref{eqn:log-likelihood} is non-negative, the first term in \eqref{eqn:log-likelihood} provides a lower bound of the log-likelihood. In fact, the expectation term in \eqref{eqn:log-likelihood} is equal to
\begin{equation}\label{elbo_ori}
	\E_{q(\rvz, \rvz_\rvn| \rvx,\rvy)}\log p(\rvx,\rvy | \rvz, \rvz_\rvn) - \KL ( q(\rvz, \rvz_\rvn| \rvx,\rvy) \| p(\rvz_\rvn, \rvz) ),
\end{equation}
which is called as the evidence lower bound (ELBO). Thus, instead of maximizing the intractable log-likelihood, we maximize the ELBO. Suppose the image information is contained in the paired data $(\rvx,\rvy)$ and the degradation information is only contained in the noisy data $\rvy$, we choose the inference model which has the decomposition
\begin{equation}\label{model:inference}
	q(\rvz_\rvn, \rvz | \rvx, \rvy) = q(\rvz | \rvx, \rvy) q(\rvz_\rvn | \rvy).
\end{equation}
Moreover, the graphical model in Figure~\ref{graph_model} gives
\begin{equation}\label{model:true}
	p(\rvz_\rvn,\rvz) = p(\rvz_\rvn)p(\rvz), \quad p(\rvx,\rvy | \rvz, \rvz_\rvn) = p(\rvx | \rvz)p(\rvy| \rvz,\rvz_\rvn).
\end{equation}
Combining \eqref{model:inference} with \eqref{model:true}, the ELBO given in \eqref{elbo_ori} satisfies
\begin{equation*}
	\begin{aligned}
		\mathrm{ELBO} =& \underbrace{\E_{q(\rvz| \rvx,\rvy)} \log p(\rvx | \rvz) + \E_{q(\rvz|\rvx,\rvy)q(\rvz_\rvn|\rvy)} \log p(\rvy | \rvz_\rvn,\rvz)}_{\mathrm{Reconstruction}} \\
		&\underbrace{ - \KL(q(\rvz | \rvx, \rvy) \| p(\rvz)) - \KL (q(\rvz_\rvn | \rvy) \| p(\rvz_\rvn))}_{\mathrm{KL}}.
	\end{aligned}
\end{equation*}
Due to the existence of $q(\rvz|\rvx,\rvy)$ in ELBO, it still needs the paired information. To further decouple this relationship, we define the \emph{inference invariant condition} as 
\begin{equation}\label{inferece_assume}
	q(\rvz| \rvx) = q(\rvz| \rvy), \quad \forall (\rvx,\rvy) \sim p(\rvx, \rvy).
\end{equation}
The above condition means that for paired data $(\rvx,\rvy)$, the latent image information $\rvz$ can be obtained from either clean image $\rvx$ or noisy image $\rvy$. In other words, the latent code $\rvz$ represents the common features between the source domain and target domain. In practice, this condition can be  satisfied using a pre-trained network or predefined operations. More discussions related to this condition is present in Section~\ref{model_details}. Under the inference invariant condition \eqref{inferece_assume} and setting the inference model $q(\rvz|\rvx,\rvy)$ to
\begin{equation}
	q(\rvz| \rvx,\rvy) := q(\rvz| \rvx) = q(\rvz| \rvy),
\end{equation}
we have ${\rm ELBO} = {\rm ELBO}_\rvx + {\rm ELBO}_\rvy$ where
\begin{align}
	\mathrm{ELBO}_{\rvx} & := \E_{q(\rvz| \rvx)} \log p(\rvx | \rvz) - \frac{1}{2} \KL  (q(\rvz | \rvx) \| p(\rvz)), \label{eqn:x} \\
	\mathrm{ELBO}_{\rvy} & := \E_{q(\rvz|\rvy)q(\rvz_\rvn|\rvy)}\log p(\rvy|\rvz_\rvn,\rvz)  \label{eqn:y} \\
	& -\frac{1}{2} \KL(q(\rvz|\rvy)||p(\rvz)) 
	- \KL (q(\rvz_\rvn | \rvy) \| p(\rvz_\rvn)). \nonumber
\end{align}
It is clear that ELBO$_{\rvx}$ is only related to clean images $\rvx$ and ELBO$_{\rvy}$ is only related to noisy images $\rvy$. Thus, no paired data is needed in the above formulation. 
In summary, the loss of the proposed LUD-VAE model is
\begin{equation*}
	\text{Total Loss} = - (\mathrm{ELBO}_{\rvx} + \mathrm{ELBO}_{\rvy}).
\end{equation*}

\begin{rem}
	The recent DeFlow model~\cite{wolf2021deflow} also estimate the joint density $p(\rvx, \rvy)$ by directly maximizing two marginal log-likelihood functions $\log p(\rvx)$ and $\log p(\rvy)$. In~\cite{wolf2021deflow}, it assumes that there are two latent variables $\rvz_\rvx$ and $\rvz_\rvy$ where $\rvz_\rvx$ and $\rvz_\rvy$ are not independent, which is the main difference with our method. If $\rvz_\rvy = \rvz_\rvx + \rvu$ where $\rvu \sim\mathcal{N}(\mu_u,\Sigma_u)$, the log-likelihood of the joint density contains marginal log-likelihood functions and one extra cross term that requires the clean-noisy image pairs. See the Appendix for the detailed derivation. It is worth mentioning that the DeFlow method introduces a domain invariant function and the conditional flow model. This modified model may help decorrelate the latent representations, but investigating the exact mechanism requires more discussion.
\end{rem}

\subsection{Method settings}\label{model_details}
In this subsection, we introduce the details of ELBO$_{\rvx}$ and ELBO$_{\rvy}$ by specifying the choices of latent variables $\rvz$, $\rvz_\rvn$ and the network architecture.

{\noindent \bf Hierarchical structure.} In ELBO$_{\rvx}$ and ELBO$_{\rvy}$, the choices of inference models $q(\rvz|\rvx), q(\rvz_\rvn|\rvy)$ and the prior distributions $p(\rvz), p(\rvz_\rvn)$ are important for the performance of our model. In classical VAE model~\cite{kingma2013auto}, it sets $p(\rvz)$ and $p(\rvz_\rvn)$ as the normal distribution that facilitates the computation of KL terms in the loss function. However, this may restrict the expressive ability of the prior distribution, and it is pointed out in~\cite{dosovitskiy2016generating} that classical VAE models usually suffer from the problem of generating unrealistic and blurry images. To address this problem, we adopt the hierarchical representations of both inference models and prior distribution~\cite{sonderby2016ladder}. Let $N$ be the number of layers, we assume 
\begin{equation*}
	\rvz=(\rvz^1,\dots,\rvz^N), \quad \rvz_\rvn=(\rvz_\rvn^1,\dots,\rvz_\rvn^N).
\end{equation*}
Thus, the prior distributions $p(\rvz)$ and $p(\rvz_\rvn)$ are
\begin{equation}\label{h_prior}
	\begin{aligned}
		&  \ p(\rvz) = p(\rvz^1) p(\rvz^2 | \rvz^1) \cdots p(\rvz^N | \rvz^{<N}), \\
		&  \ p(\rvz_\rvn) = p(\rvz_\rvn^1) p(\rvz_\rvn^2 | \rvz_\rvn^1) \cdots p(\rvz_\rvn^N | \rvz_\rvn^{<N}),
	\end{aligned}
\end{equation}
where $\rvz^{<K} = (\rvz^1,\dots,\rvz^{K-1})$ and $\rvz_\rvn^{<K} = (\rvz_\rvn^1,\dots,\rvz_\rvn^{K-1})$. 
Similarly, we impose the same order on the inference models to generate latent variables:
\begin{equation}\label{h_posterior}
	\begin{aligned}
		& q(\rvz|\rvx) = q(\rvz^1|\rvx) q(\rvz^2|\rvz^1,\rvx) \cdots q(\rvz^N | \rvz^{<N}, \rvx), \\
		& q(\rvz_\rvn | \rvy) = q(\rvz_\rvn^1 |\rvy) q(\rvz_\rvn^2 | \rvz_\rvn^1,\rvy) \cdots q(\rvz_\rvn^N | \rvz_\rvn^{<N},\rvy).
	\end{aligned}
\end{equation}
Combining~\eqref{h_prior} and~\eqref{h_posterior}, the KL divergence in ELBO$_{\rvx}$ is 
\begin{equation}\label{KL:factorized_x}
	\begin{aligned}
		\KL & (q(\rvz  | \rvx)   \| p(\rvz)) =  \KL(q(\rvz^1|\rvx)\|p(\rvz^1)) \\
		&+ \sum_{l=2}^{N} \E_{q(\rvz^{<l}|\rvx)} \left[\KL(q(\rvz^l|\rvz^{<l},\rvx) \| p(\rvz^l|\rvz^{<l})) \right],
	\end{aligned}
\end{equation}
and the two KL terms in ELBO$_{\rvy}$ are 
\begin{align}
	\KL & (q(\rvz  | \rvy)   \| p(\rvz)) =  \KL(q(\rvz^1|\rvy)\|p(\rvz^1)) \nonumber\\
	&+ \sum_{l=2}^{N} \E_{q(\rvz^{<l}|\rvy)} \left[\KL(q(\rvz^l|\rvz^{<l},\rvy) \| p(\rvz^l|\rvz^{<l})) \right], \label{KL:factorized_y1}\\
	\KL & (q(\rvz_\rvn  | \rvy)   \| p(\rvz_\rvn)) =  \KL(q(\rvz_\rvn^1|\rvy)\|p(\rvz_\rvn^1)) \nonumber\\
	&+ \sum_{l=2}^{N} \E_{q(\rvz_\rvn^{<l}|\rvy)} \left[\KL(q(\rvz_\rvn^l|\rvz_\rvn^{<l},\rvy) \| p(\rvz_\rvn^l|\rvz_\rvn^{<l})) \right]. \label{KL:factorized_y2}
\end{align}
\begin{rem}
The hierarchical structures on latent variables~$\rvz$ and $\rvz_\rvn$ impose the conditional dependences of for different layers. Compared to the classical VAE method, it significantly improves the expressive ability of the prior distributions for modeling the complex degradation in image processing.
\end{rem}

{\noindent \bf The choice of latent space.} Besides the hierarchical structure on latent variables~$\rvz_\rvn$, $\rvz$, we specify their distributions for facilitating the computation. For $p(\rvz_\rvn)$, we assume
\begin{equation*}
	\begin{aligned}
		& p(\rvz_n^1) = \gN (\mathbf{0}, \rmI), \quad p(\rvz_\rvn^l|\rvz_\rvn^{<l}) = \gN (\mu^l_{p}(\rvz_\rvn^{<l}), \sigma^l_{p}(\rvz_\rvn^{<l})), \\
		& q(\rvz_\rvn^l|\rvz_\rvn^{<l},\rvy) = \gN (\mu^l_{q}(\rvz_\rvn^{<l}, \rvy), \sigma^l_{q}(\rvz_\rvn^{<l}, \rvy) ), 
	\end{aligned}
\end{equation*}
where $\mu^l_{p}$, $\sigma^l_{p}$, $\mu^l_{q}$, $\sigma^l_{q}$ are encoding neural networks for the degradation part. Thus, the KL divergence in~\eqref{KL:factorized_y2} is computable. For $p(\rvz)$, we assume
\begin{equation*}
	\begin{aligned}
		& p(\rvz^1) = \gU(\Omega^1), \quad p(\rvz^l|\rvz^{<l}) = \gU(\Omega^l), \\
		& q(\rvz^l|\rvz^{<l},\rvx) = \delta (e^l(\rvz^{<l},\rvx)), \quad q(\rvz^l|\rvz^{<l},\rvy) = \delta (e^l(\rvz^{<l},\rvy)),
	\end{aligned}
\end{equation*}
where $e^l$ is the encoding neural network for the image contents, $\Omega^l$ is the range of network $e^l$, $\gU(\Omega^l)$ denotes the uniform distribution on $\Omega^l$, $\delta(\cdot)$ denotes the delta distribution. In this case, we have
\begin{equation}
	\KL(q(\rvz | \rvx) \| p(\rvz)) = c, \quad \KL (q(\rvz | \rvy) \| p(\rvz)) = c,
\end{equation}
where $c$ is a constant. Moreover, let $d(\cdot)$ be the decoding neural network and assume 

\begin{equation*}
	\begin{aligned}
		& p(\rvx|\rvz) = p(\rvx|\rvz^N) = \gN(d(\rvz^N), \rmI), \\
		& \rvd(\rvx) = d(\rvz^N), \quad \rvz^l = e^l(\rvz^{<l},\rvx),
	\end{aligned}
\end{equation*}
the reconstruction term in ELBO$_{\rvx}$ becomes
\begin{equation}\label{eqn:expectation_x}
	\E_{q(\rvz | \rvx)} \log p(\rvx | \rvz) = - \frac{1}{2} \| \rvd(\rvx) - \rvx \|^2.
\end{equation}
Similarly, the reconstruction term in ELBO$_{\rvy}$ is
\begin{equation}\label{eqn:expectation_y}
	\E_{q(\rvz|\rvy)q(\rvz_\rvn|\rvy)}\log p(\rvy|\rvz_\rvn,\rvz) = - \frac{1}{2} \E_{q(\rvz_\rvn|\rvy)} \| \rvd(\rvy) - \rvy \|^2,
\end{equation}
where we assume 
\begin{equation*}
	\begin{aligned}
		& p(\rvy|\rvz, \rvz_\rvn) = p(\rvy|\rvz^N, \rvz_\rvn^N) = \gN(d(\rvz^N,\rvz_\rvn^N), \rmI), \\
		& \rvd(\rvy) = d(\rvz^N,\rvz_\rvn^N), \quad \rvz^l = e^l(\rvz^{<l},\rvy), \\
		& \rvz_\rvn^l \sim \gN (\mu^l_{q}(\rvz_\rvn^{<l}, \rvy), \sigma^l_{q}(\rvz_\rvn^{<l}, \rvy)).
	\end{aligned}
\end{equation*}
The reparametrization technique~\cite{kingma2013auto} can be used for estimating~\eqref{eqn:expectation_y}. In this work, we set the number of layers $N=3$. See Figure~\ref{model_arch_fig} for the details of the network architecture. The basic convolution block contains four $3 \times 3$ convolution layers with Rectified Linear Unit (ReLU) activation function. The encoding function $\mu^l_{p}$, $\sigma^l_{p}$, $\mu^l_{q}$, $\sigma^l_{q}$, and $e^l$ are chosen as a $1 \times 1$ convolution layer. The inference and generative processes share weights at red convolution blocks.

\begin{rem}
To synthetic the paired training data, it is sufficient to generate the degradation of clean images rather than the image content. Thus, we model the image part as an auto-encoder without generation ability and the degradation part as the hierarchical VAE model~\cite{vahdat2020nvae,child2020very}. This choice can strengthen the reconstruction ability of our model. A similar idea has been proposed in the VQ-VAE model~\cite{oord2017neural,razavi2019generating}, which uses a discrete latent space and shows excellent generative ability. However, in the VQ-VAE model, it needs to construct a discrete codebook for $p(\rvz)$ and sample with a PixelCNN~\cite{oord2016conditional} model that increases the training difficulty. In our model, since there are no constraints on the inference model $q(\rvz|\cdot)$ for the image part, we do not need to construct a codebook for latent variable~$\rvz$. 
\end{rem}

{\noindent \bf Inference invariant condition.} The inference invariant condition \eqref{inferece_assume} is a key tool for disentangling the image part and degradation part. In fact, it assumes that for the paired clean and noisy data $(\rvx, \rvy)\sim p(\rvx, \rvy)$, the inferred latent variable $\rvz$ is the same. This condition holds if we can extract the common information between $\rvx$ and $\rvy$. In our method, we pass the input data through a pre-processing operator~$h$, which can be manually designed or trained by an extra dataset. In LUD-VAE, the pre-processor $h$ adds the Gaussian noise into clean images and noisy images. That is, given arbitrary clean image $\rvx_i$ and noisy image $\rvy_j$, we define
\begin{equation}\label{iic}
	h(\rvx_i) = \rvx_i + \rvn_\rvx, \quad h(\rvy_j) = \rvy_j + \rvn_\rvy,
\end{equation}
where $\rvn_\rvx \sim \gN(\mathbf{0}, \sigma_\rvx^2 \rmI)$ and $\rvn_\rvy \sim \gN(\mathbf{0}, \sigma_\rvy^2 \rmI)$. In the next, we give explanations that $h$ is a reasonable pre-processor for the task of Gaussian noise removal.

Assume $(\rvx,\rvy)$ is the paired data and $\rvy = \rvx + \rvn$, where $\rvn\sim\gN(\mathbf{0},\sigma^2\rmI)$, we choose $q(\rvz|\rvx,\rvn_\rvx) = q(\rvz|\rvx + \rvn_\rvx)$ and $q(\rvz|\rvy,\tilde{\rvn}_\rvy) = q(\rvz|\rvx + \tilde{\rvn}_\rvy)$. Thus, the inference model becomes
\begin{equation}\label{eqn:q}
	\begin{aligned}
		& q(\rvz | \rvx) = \E_{p(\rvn_\rvx)} q(\rvz | \rvx + \rvn_\rvx), \\
		& q(\rvz | \rvy) = \E_{p(\tilde{\rvn}_\rvy)} q(\rvz | \rvx + \tilde{\rvn}_\rvy),
	\end{aligned}
\end{equation}
where $\tilde{\rvn}_\rvy = \rvn + \rvn_\rvy \sim \gN(\rvn, \sigma_\rvy^2 \rmI)$. Thus, the difference between $q(\rvz | \rvx)$ and $q(\rvz | \rvy)$ comes from the distance between $p(\rvn_\rvx)$ and $p(\tilde{\rvn}_\rvy)$, and it has
\begin{equation}
    \KL(p(\tilde{\rvn}_\rvy) \| p(\rvn_\rvx) )  = \log \frac{ \sigma^{K}_{\rvx} }{ \sigma^{K}_{\rvy} } - \frac{K}{2} +  \frac{K\sigma_\rvy^2}{2\sigma_\rvx^2} + \frac{\|\rvn \|_2^2}{2\sigma_\rvx^2},
\end{equation}
where $K$ is the dimension of random variable $\rvn_\rvx$. Let $\sigma_\rvx$, $\sigma_\rvy$ go to infinity and $\lim_{\sigma_\rvx,\sigma_\rvy\to\infty}{\sigma_\rvx/\sigma_\rvy} = 1$, the above KL divergence approaches to $0$. Thus, the inference invariant condition holds in asymptotic sense. See more details in Appendix and the quantitative results Table~\ref{aim_iic} with different noise levels.

\begin{rem}
In this work, we use Monte Carlo method to estimate the above expectation terms in \eqref{eqn:q} and the number of sampling is set to 1 for each iterations.
\end{rem}

\begin{figure}
	\centering
	\begin{tabular}{c@{\hspace{0.01\linewidth}}c@{\hspace{0.01\linewidth}}c}
		\includegraphics[width=0.32 \linewidth]{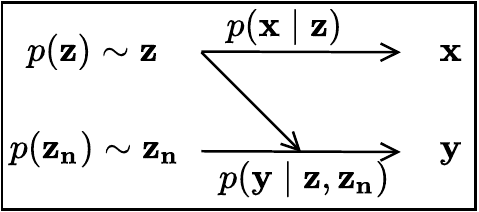} &
		\includegraphics[width=0.32 \linewidth]{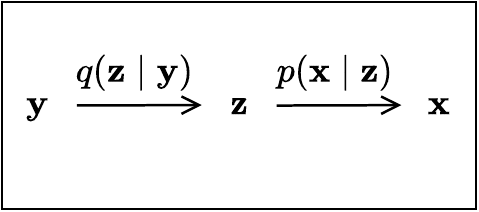} &
		\includegraphics[width=0.32 \linewidth]{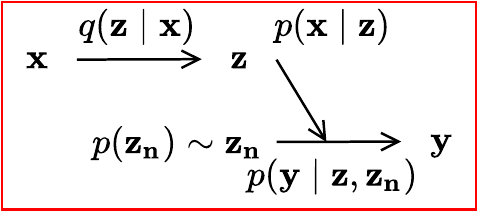} \\
		\small{L2CN} & \small{N2C} & \small{C2N} \\
	\end{tabular}
	\caption{Three methods to generate paired data. C2N is used in LUD-VAE.}
	\label{three_methods}
\end{figure}

{\noindent \bf Generate synthetic paired data.} After the training procedure, there are three methods to generate paired data.
\begin{enumerate}
	\item L2CN: Latent to clean and noisy pair. Sample $\rvz_\rvn\sim p(\rvz_\rvn)$, $\rvz \sim p(\rvz)$, then generate $(\rvx, \rvy)$ from $p(\rvy | \rvz, \rvz_\rvn)$ and $p(\rvx | \rvz)$.
	\item N2C: Noisy to clean. Sample $\rvy \sim p(\rvy)$, inference the latent variable $\rvz$ with $q(\rvz | \rvy)$, then generate the corresponding $\rvx$ from $p(\rvx | \rvz)$.
	\item C2N: Clean to noisy. Sample $\rvx \sim p(\rvx)$, inference the latent variable $\rvz$ with $q(\rvz | \rvx)$, sample $\rvz_\rvn\sim p(\rvz_\rvn)$, then generate the corresponding $\rvy$ from $p(\rvy | \rvz, \rvz_\rvn)$.
\end{enumerate}
See Figure~\ref{three_methods} for the graphical explanations, and we adopt C2N to generate paired training data for the downstream tasks. This choice is consistent with the choice of the space of latent variable $\rvz$ since it does not have the generation ability. Moreover, the results in Table~\ref{sidd_m2_m3} show that C2N achieves better performance than the N2C method.

\begin{table*}
	\setlength{\tabcolsep}{15pt}
	\renewcommand{\arraystretch}{1.3}
	\centering
	\caption{Quantitative comparison on real-world super-resolution datasets AIM19 and NTIRE20. The synthetic datasets are trained with the ESRGAN model~\cite{wang2018esrgan}. $^*$ denotes the results are taken from their original papers.}
	\begin{tabular}{ccccccc}
		\hline
		\specialrule{0em}{.9pt}{.9pt}
		\rowcolor[gray]{0.95} 
		\cellcolor[gray]{0.95} & \multicolumn{3}{c}{\cellcolor[gray]{0.95}AIM19}                                                            & \multicolumn{3}{c}{\cellcolor[gray]{0.95}NTIRE20} \\
		\rowcolor[gray]{0.95} 
		\multirow{-2}{*}{\cellcolor[gray]{0.95}Method} & PSNR $\uparrow$  & SSIM $\uparrow$ & LPIPS $\downarrow$                                                    & PSNR $\uparrow$  & SSIM $\uparrow$ & LPIPS $\downarrow$ \\
		\specialrule{0em}{.9pt}{.9pt}
		\hline
		ESRGAN-Bicubic                                      & 21.69 & 0.5517 & 0.517 & 20.45 & 0.3241 & 0.675 \\
		FSSR~\cite{fritsche2019frequency}                   & 20.81 & 0.5242 & 0.387 & 21.07 & 0.4356 & 0.414 \\
		Impressionism~\cite{ji2020real}                     & 21.99 & 0.6060 & 0.420 & 25.27 & 0.6731 & 0.229 \\
		DASR~\cite{wei2021unsupervised}                     & 21.06 & 0.5658 & 0.375 & 23.70 & 0.5748 & 0.328 \\
		DeFlow-NP~\cite{wolf2021deflow}                     & 21.06 & 0.5842 & 0.346 & 24.81 & 0.6777 & 0.225 \\
		\textbf{LUD-VAE} (ours)                             & \textbf{22.32} & \textbf{0.6197} & \textbf{0.341} & \textbf{25.79} & \textbf{0.7178} & \textbf{0.219} \\ \hline
		DeFlow~\cite{wolf2021deflow}                        & 22.25 & 0.6214 & 0.349 & 25.87 & 0.7005 & 0.218 \\
		CinCGAN~\cite{yuan2018unsupervised}                 & 21.60 & 0.6129 & 0.461 & 24.83 & 0.6752 & 0.509 \\ 
		Wang {\it et al.}$^*$~\cite{wang2021unsupervised}   & 22.60 & 0.622  & 0.340 & 25.40 & 0.707  & 0.252 \\ 
		Yoon {\it et al.}$^*$~\cite{yoon2021simple}         & -     & -      & -     & 24.30 & 0.6731 & 0.282 \\ 
		\hline
	\end{tabular}
	\label{aim_ntire}
\end{table*}

\section{Experiments and results}
We evaluate the performance of our LUD-VAE method on real-world image denoising, super-resolution, and low-light image enhancement tasks. First, we use LUD-VAE to learn the unknown degradation model under the unpaired learning settings, generate the synthetic training dataset, and then use an off-the-shelf supervised learning algorithm to learn the restoration model. All experiments are evaluated in the sRGB space.

\subsection{Datasets and evaluation metrics}
Two real-world super-resolution datasets, two real-world denoising datasets, and one low-light image enhancement dataset are chosen to evaluate our method:

{\noindent \bf AIM19}: Track 2 of the AIM 2019 real-world super-resolution challenge~\cite{lugmayr2019aim} provides a dataset of unpaired noisy-clean images. The noisy images are synthesized with an unknown combination of noise and compression, which mainly behave as structural and low frequency noises. The task is to learn a super-resolution model from the unpaired dataset, which restores high-resolution clean images from the low-resolution noisy inputs. The challenge also provides a validation set of 100 paired images, where different models can be compared with quantitative metrics. We refer to this dataset as the AIM19 dataset.

{\noindent \bf NTIRE20}: Track 1 of NTIRE 2020 super-resolution challenge~\cite{lugmayr2020ntire} follows the same setting as the AIM19 dataset, where it features an entirely different type of degradation, namely highly correlated high-frequency noise. As AIM19, there is a validation set containing 100 paired images that can be quantitatively evaluated. We refer to this dataset as the NTIRE20 dataset.

{\noindent \bf SIDD}: The smartphone image denoising dataset (SIDD)~\cite{abdelhamed2018high} provides 30,000 noisy images from 10 scenes under different lighting conditions using five representative smartphone cameras and generates their ground truth images. We use the SIDD-Small dataset and ignore original indexes of clean noisy images to set up an unpaired dataset. It also provides the validation and benchmark datasets, each of which is cropped into 32 blocks of size $256 \times 256$, resulting in 1280 image blocks in each dataset. We refer to this dataset as the SIDD dataset.

{\noindent \bf DND}: The Darmstadt Noise Dataset dataset (DND)~\cite{plotz2017benchmarking} contains 50 real-world noisy images taken by four commercial cameras. We use the clean images from the SIDD-Small dataset and DND's noisy images to set up the unpaired dataset. DND provides the benchmark dataset containing 1000 $512 \times 512$ image blocks. We refer to this dataset as the DND dataset.

{\noindent \bf LOL}: The Low-light dataset (LOL)~\cite{wei2018deep} contains 485 paired images for training and 15 paired images for testing. Since our model is trained with unpaired data, we use the unpaired training dataset provided by EnlightenGAN~\cite{jiang2021enlightengan}, which contains 914 low-light and  1016 normal-light images. There is no intersection between EnlightenGAN's training data and LOL's testing data, so we train our model on EnlightenGAN's unpaired dataset and evaluate it on LOL's testing dataset. We refer to this dataset as the LOL dataset.

\begin{table}
	\setlength{\tabcolsep}{8pt}
	\renewcommand{\arraystretch}{1.3}
	\centering
	\caption{Synthetic degradation comparison on real-world super-resolution datasets AIM19 and NTIRE20. The top two methods are indicated with color red and blue, respectively.}
	\begin{tabular}{ccccc}
		\hline
		\specialrule{0em}{.9pt}{.9pt}
		\rowcolor[gray]{0.95} 
		\cellcolor[gray]{0.95} & \multicolumn{2}{c}{\cellcolor[gray]{0.95}AIM19}                                                            & \multicolumn{2}{c}{\cellcolor[gray]{0.95}NTIRE20} \\
		\rowcolor[gray]{0.95} 
		\multirow{-2}{*}{\cellcolor[gray]{0.95}Method}  & AKLD $\downarrow$ & FID $\downarrow$ & AKLD $\downarrow$ & FID $\downarrow$ \\
		\specialrule{0em}{.9pt}{.9pt}
		\hline
		Bicubic                           & 0.701 & 112.9 & 0.701 & 56.3  \\
		FSSR~\cite{fritsche2019frequency} & 0.379 & {\color{blue} 67.6} & 0.224 & 40.2  \\
		Impressionism~\cite{ji2020real}   & 0.549 & 87.3 & 0.275 & {\color{blue} 26.5}  \\
		DASR~\cite{wei2021unsupervised}   & {\color{red} 0.328} & 72.4 & 0.346 & 48.6  \\
		DeFlow~\cite{wolf2021deflow}      & 0.356 & 119.8 & {\color{blue} 0.156} & 34.7  \\
		\textbf{LUD-VAE} (ours)           & {\color{blue} 0.329} & {\color{red} 57.2} & {\color{red} 0.120} & {\color{red} 24.9}  \\ \hline
	\end{tabular}
	\label{noise_compare}
\end{table}

For all five datasets, we report the peak signal-to-noise ratio (PSNR) and the structural similarity index (SSIM)~\cite{wang2004image}. For AIM19, NTIRE20, and LOL datasets, we also compute the LPIPS~\cite{zhang2018unreasonable} distance, which is based on the comparison between features of a neural network. Here we use a pre-trained AlexNet~\cite{krizhevsky2012imagenet} model.

\begin{figure*}
	\centering
	\begin{tabular}{c@{\hspace{0.005\linewidth}}c@{\hspace{0.005\linewidth}}c@{\hspace{0.005\linewidth}}c@{\hspace{0.005\linewidth}}c@{\hspace{0.005\linewidth}}c@{\hspace{0.005\linewidth}}c}
		\includegraphics[width=0.135\linewidth]{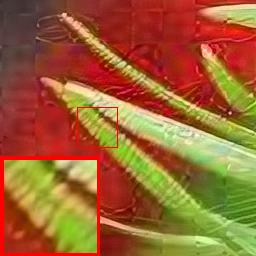}   &
		\includegraphics[width=0.135\linewidth]{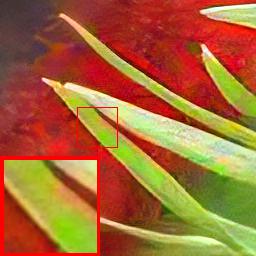}   &
		\includegraphics[width=0.135\linewidth]{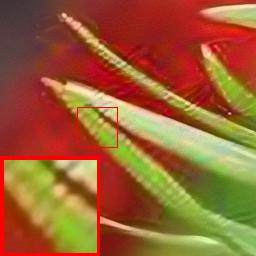} &
		\includegraphics[width=0.135\linewidth]{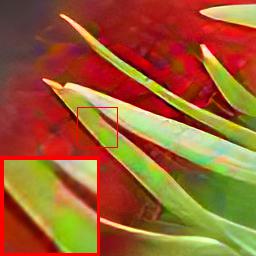} &
		\includegraphics[width=0.135\linewidth]{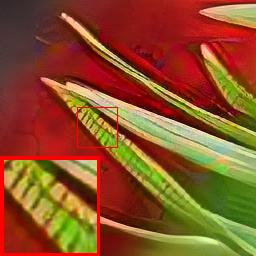} &
		\includegraphics[width=0.135\linewidth]{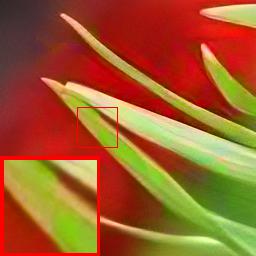} &
		\includegraphics[width=0.135\linewidth]{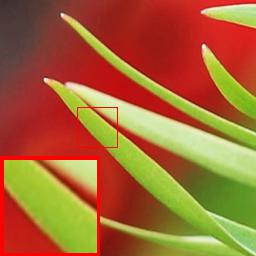} \\
		
		\includegraphics[width=0.135\linewidth]{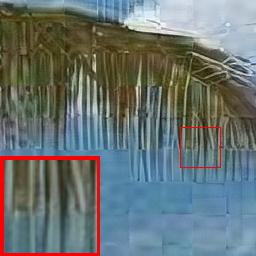}   &
		\includegraphics[width=0.135\linewidth]{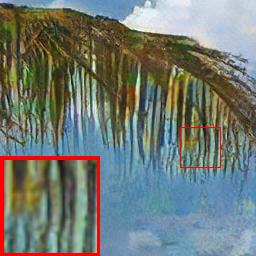}   &
		\includegraphics[width=0.135\linewidth]{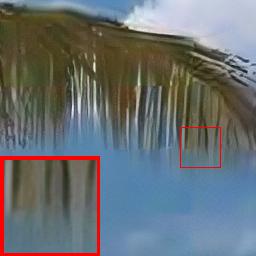} &
		\includegraphics[width=0.135\linewidth]{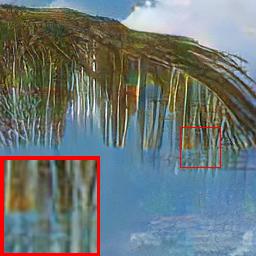} &
		\includegraphics[width=0.135\linewidth]{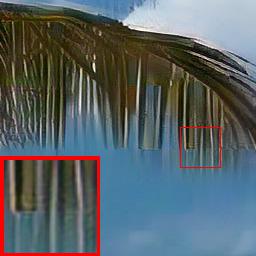} &
		\includegraphics[width=0.135\linewidth]{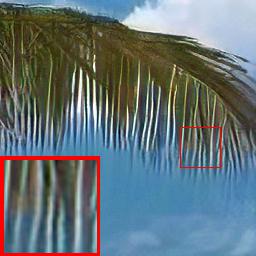} &
		\includegraphics[width=0.135\linewidth]{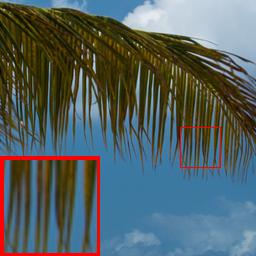} \\
		
		\includegraphics[width=0.135\linewidth]{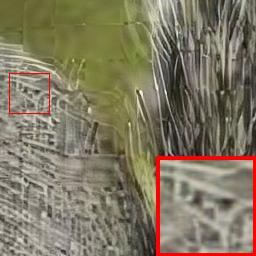}   &
		\includegraphics[width=0.135\linewidth]{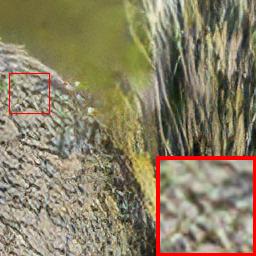}   &
		\includegraphics[width=0.135\linewidth]{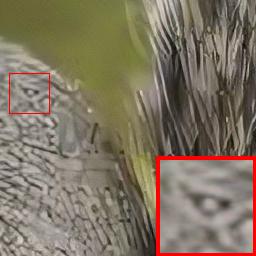} &
		\includegraphics[width=0.135\linewidth]{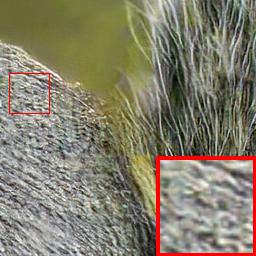} &
		\includegraphics[width=0.135\linewidth]{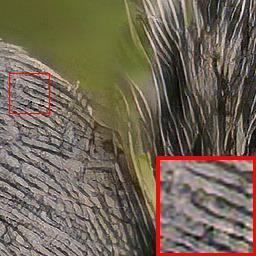} &
		\includegraphics[width=0.135\linewidth]{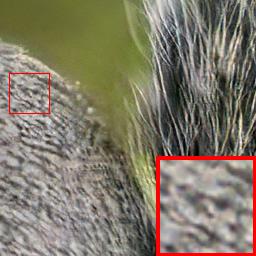} &
		\includegraphics[width=0.135\linewidth]{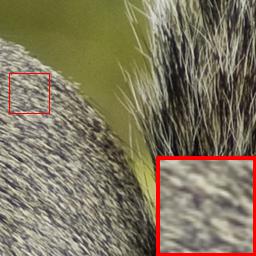} \\
		
		\includegraphics[width=0.135\linewidth]{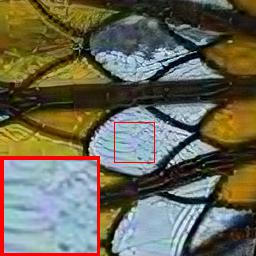}   &
		\includegraphics[width=0.135\linewidth]{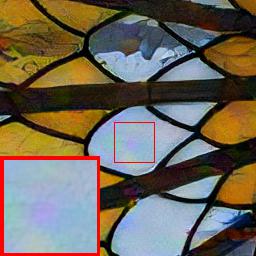}   &
		\includegraphics[width=0.135\linewidth]{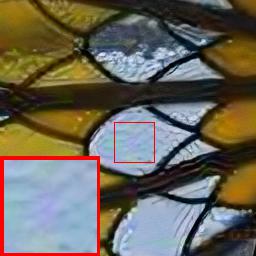} &
		\includegraphics[width=0.135\linewidth]{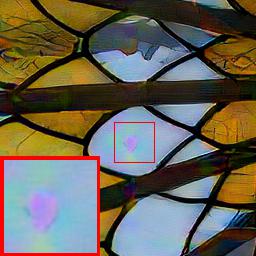} &
		\includegraphics[width=0.135\linewidth]{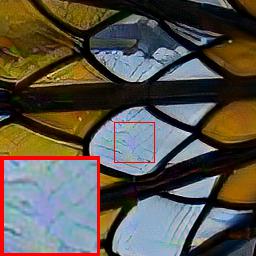} &
		\includegraphics[width=0.135\linewidth]{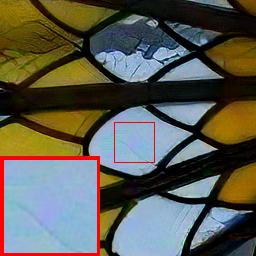} &
		\includegraphics[width=0.135\linewidth]{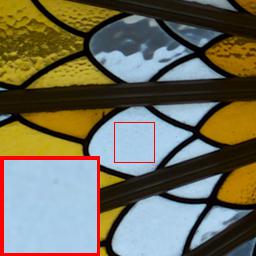} \\
		
		\includegraphics[width=0.135\linewidth]{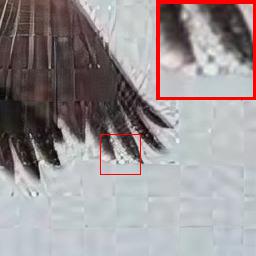}   &
		\includegraphics[width=0.135\linewidth]{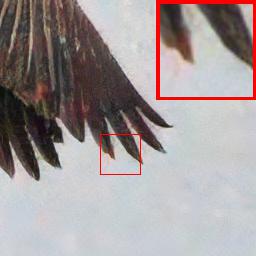}   &
		\includegraphics[width=0.135\linewidth]{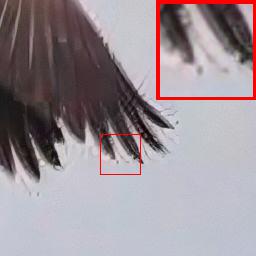} &
		\includegraphics[width=0.135\linewidth]{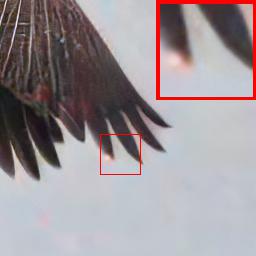} &
		\includegraphics[width=0.135\linewidth]{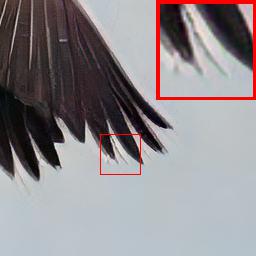} &
		\includegraphics[width=0.135\linewidth]{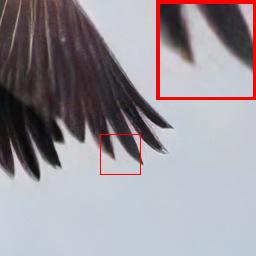} &
		\includegraphics[width=0.135\linewidth]{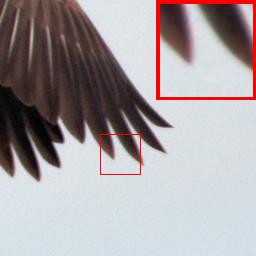} \\
		
		\small{ESRGAN-Bicubic} & \small{FSSR} & \small{Impressionism} & \small{DASR} & \small{DeFlow-NP} & \small{LUD-VAE} & \small{Ground Truth}
	\end{tabular}
	\caption{Visual comparison on real-world super-resolution dataset AIM19.}
	\label{rwsr_aim}
\end{figure*}

\subsection{Implementation details}

We train all LUD-VAE models for 200k iterations with the Adam~\cite{kingma2014adam} optimizer. The initial learning rate is set to $1e-4$ and is halved after 100k iterations. We use a batch size of 16, containing random crops of size $64 \times 64$. Batches are sampled randomly so that images from each domain are drawn with the same possibility. Random flips and rotates are used as data augmentation. Besides, to avoid the posterior collapse, we apply the KL annealing method~\cite{bowman2015generating}, and use the linear anneal scheme for $\KL (q(\rvz_\rvn | \rvy_i) \| p(\rvz_\rvn))$ in the first 10k iteration.

\begin{figure}
	\centering
	\includegraphics[width=1.\linewidth]{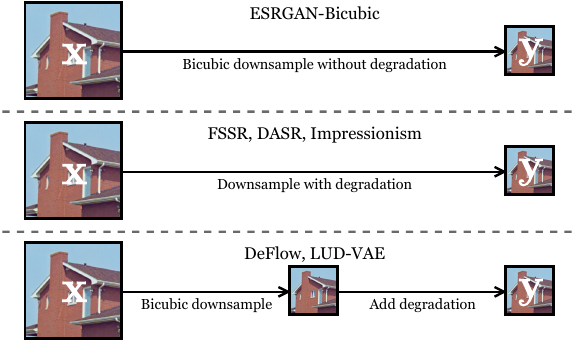}
	\caption{The synthesis of training data in super-resolution.}
	\label{sr_gen}
\end{figure}

{\noindent \bf Super-resolution case.} For real-world super-resolution, the degradation process is
\begin{equation}
	\rvy = \gD(\rvx) + \rvn,
\end{equation}
where $\gD$ is an unknown downsample operator, and $\rvn$ is an unknown noise. To simplify this problem, we replace $\gD$ by the bicubic downsample $\gB$ and put the approximation error of $\gB$ and $\gD$ into $\rvn$, \textit{i.e.,}
\begin{equation}
	\rvy = \gB(\rvx) + \rvn^{\prime},
\end{equation}
where $\rvn^{\prime} = \rvn + \gD(\rvx) - \gB(\rvx)$. Then for the given unpaired high-resolution clean dataset $\{\rvx_{i}\}$ and low-resolution noisy dataset $\{\rvy_{j}\}$, we first bicubic downsample $\{\rvx_{i}\}$ to low-resolution clean dataset $\{\gB(\rvx_{i})\}$, and then use LUD-VAE to learn the noise $\rvn^{\prime}$ from $\{\gB(\rvx_{i})\}$ and $\{\rvy_{j}\}$. After the training process, we use LUD-VAE to transfer the low-resolution clean dataset $\{\gB(\rvx_{i})\}$ to the synthetic degraded dataset, forming a paired training set with the high-resolution clean dataset $\{\rvx_{i}\}$. For the AIM19 dataset, we normalize the noisy image to make it has the same channel-wise mean and standard deviation as the clean domain, and then de-normalize the synthetic noisy image to constitute the training dataset. In AIM19, we set the $\sigma_\rvx=15$, $\sigma_\rvy=10$ in \eqref{iic} for AIM19 and $\sigma_\rvx=8$, $\sigma_\rvy=3$ for NTIRE20.

{\noindent \bf Denoising case.} For the denoising case, the degradation process becomes
\begin{equation}
	\rvy = \rvx + \rvn.
\end{equation}
We use LUD-VAE to learn the noise $\rvn$ from unpaired clean $\{\rvx_{i}\}$ and noisy $\{\rvy_{j}\}$ datasets directly. After training, we transform the clean image into the corresponding noisy image to obtain the paired training set. We set $\sigma_\rvx=40$, $\sigma_\rvy=20$ in \eqref{iic} for SIDD dataset and $\sigma_\rvx=20$, $\sigma_\rvy=10$ for DND dataset.

{\noindent \bf Low-light image enhancement case.} For the low-light image enhancement problem, we assume the degradation process as the gamma correction model:
\begin{equation}
    \rvy = c \rvx^{\gamma} + \rvn, 
\end{equation}
where $\rvx$ and $\rvy$ are in the range $[0,1]$. For the given unpaired normal-light clean image $\{ \rvx_i \}$ and low-light noisy image $\{ \rvy_j \}$, we use $\{ c_i \rvx_{i}^{\gamma_i} \}$ and $\{ \rvy_j \}$ to be the unpaired training dataset for our model, where $c_i$ and $\gamma_i$ are uniformly sampled from $[0.15, 0.4]$ and $[1.5, 2.5]$, respectively. In LOL dataset, we set $\sigma_\rvx=3$, $\sigma_\rvy=2$ in \eqref{iic}.

\subsection{Comparison of synthetic degradation.}
We compute the Average KL Divergence (AKLD)~\cite{yue2020dual} and Fréchet Inception Distance~\cite{heusel2017gans} (FID) to compare the quality of generated degradation images between different methods. AKLD computes the KL divergence between the synthesized and real noise distributions, and FID is a metric that calculates the distance between the real and the generated images, where a smaller score means better generation quality. In practice, we centrally crop the images to $128 \times 128$ from the validation set of AIM19 and NTIRE20 datasets and compare the MMD and FID metrics of different degradation modeling methods between the real and synthetic degradation images on these cropped images. The results are shown in Table~\ref{noise_compare}. From Table~\ref{noise_compare}, our method achieves the best FID on AIM19 and NTIRE20, the second best AKLD on AIM19, and the best AKLD on NTIRE 20. It is worth mentioning that finding the appropriate metric for the synthetic noisy images is difficult and deserves further exploration since both AKLD and FID metrics cannot fully reflect the performance of the downstream image restoration algorithm. For example, the DeFlow~\cite{wolf2021deflow} model has a poor FID score on the AIM19 dataset, but the restoration result outperforms other GAN-based models. 

\begin{figure*}
	\centering
	\begin{tabular}{c@{\hspace{0.005\linewidth}}c@{\hspace{0.005\linewidth}}c@{\hspace{0.005\linewidth}}c@{\hspace{0.005\linewidth}}c@{\hspace{0.005\linewidth}}c@{\hspace{0.005\linewidth}}c}
		\includegraphics[width=0.135\linewidth]{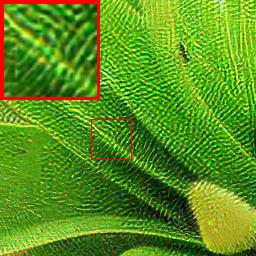}   &
		\includegraphics[width=0.135\linewidth]{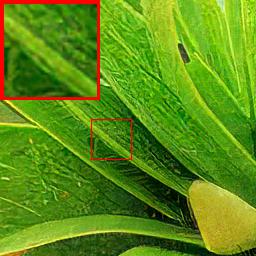}   &
		\includegraphics[width=0.135\linewidth]{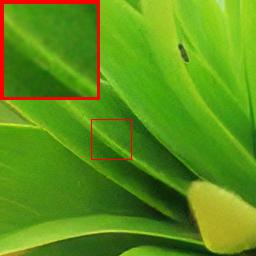} &
		\includegraphics[width=0.135\linewidth]{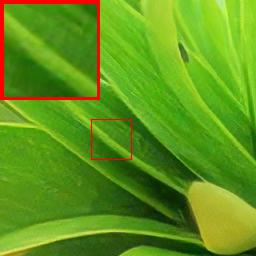} &
		\includegraphics[width=0.135\linewidth]{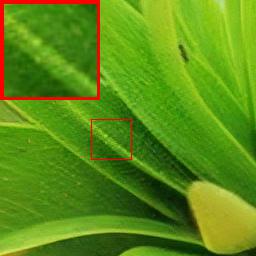} &
		\includegraphics[width=0.135\linewidth]{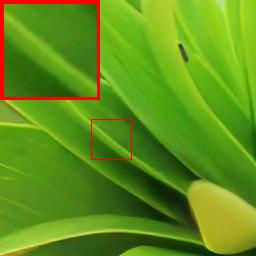} &
		\includegraphics[width=0.135\linewidth]{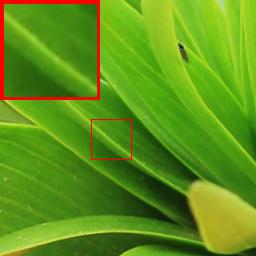} \\
		
		\includegraphics[width=0.135\linewidth]{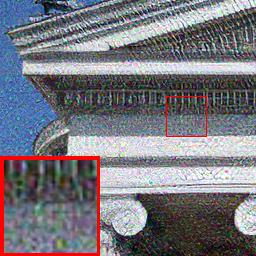}   &
		\includegraphics[width=0.135\linewidth]{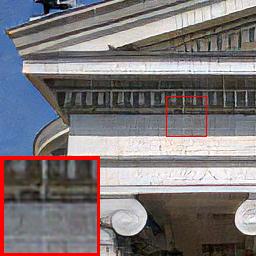}   &
		\includegraphics[width=0.135\linewidth]{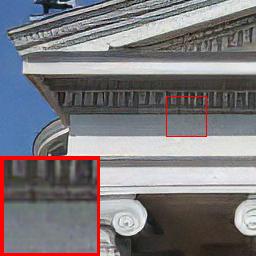} &
		\includegraphics[width=0.135\linewidth]{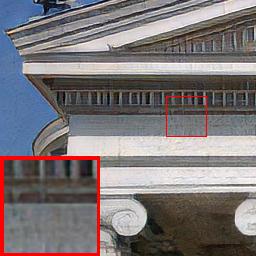} &
		\includegraphics[width=0.135\linewidth]{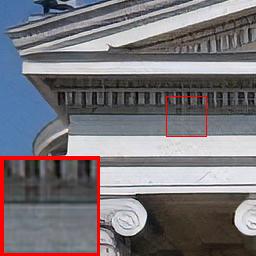} &
		\includegraphics[width=0.135\linewidth]{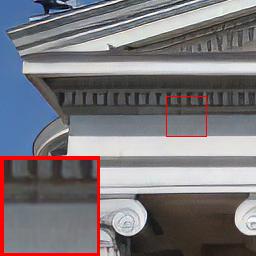} &
		\includegraphics[width=0.135\linewidth]{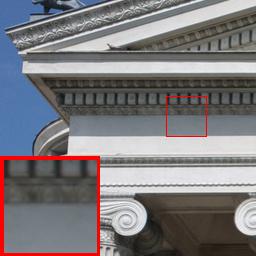} \\
		
		\includegraphics[width=0.135\linewidth]{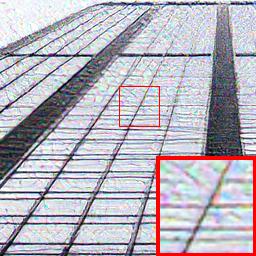}   &
		\includegraphics[width=0.135\linewidth]{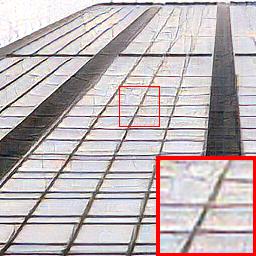}   &
		\includegraphics[width=0.135\linewidth]{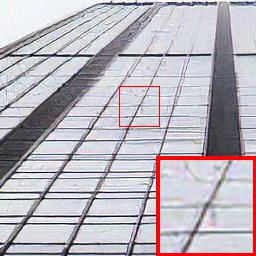} &
		\includegraphics[width=0.135\linewidth]{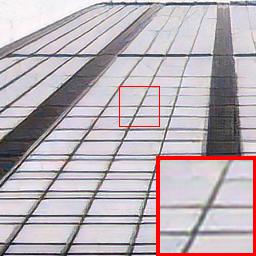} &
		\includegraphics[width=0.135\linewidth]{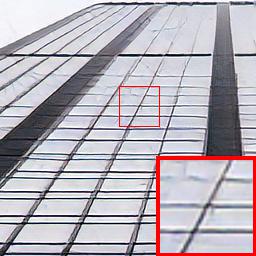} &
		\includegraphics[width=0.135\linewidth]{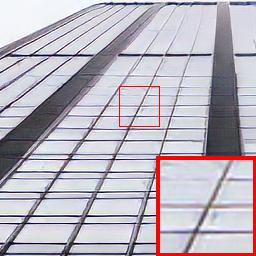} &
		\includegraphics[width=0.135\linewidth]{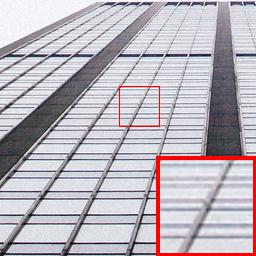} \\
		
		\includegraphics[width=0.135\linewidth]{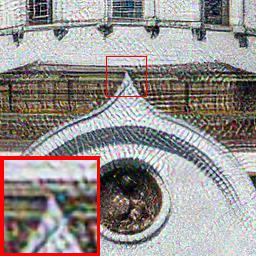}   &
		\includegraphics[width=0.135\linewidth]{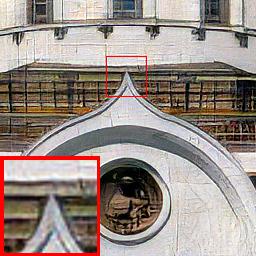}   &
		\includegraphics[width=0.135\linewidth]{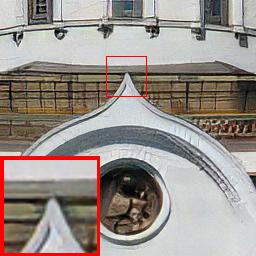} &
		\includegraphics[width=0.135\linewidth]{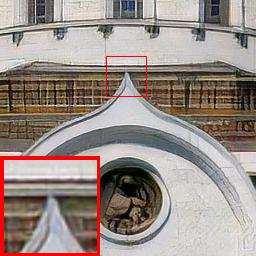} &
		\includegraphics[width=0.135\linewidth]{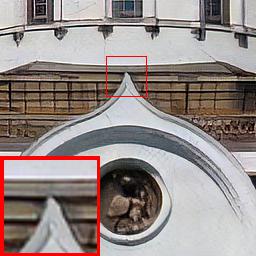} &
		\includegraphics[width=0.135\linewidth]{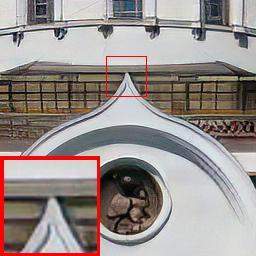} &
		\includegraphics[width=0.135\linewidth]{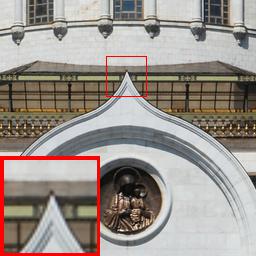} \\
		
		\includegraphics[width=0.135\linewidth]{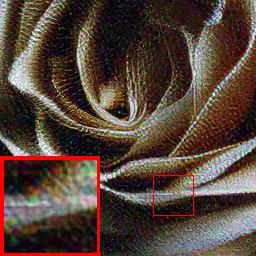}   &
		\includegraphics[width=0.135\linewidth]{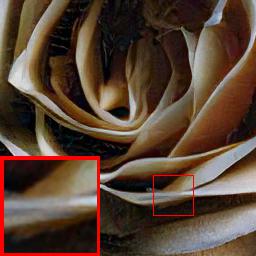}   &
		\includegraphics[width=0.135\linewidth]{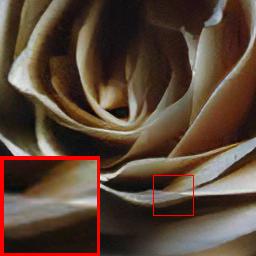} &
		\includegraphics[width=0.135\linewidth]{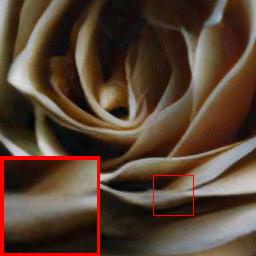} &
		\includegraphics[width=0.135\linewidth]{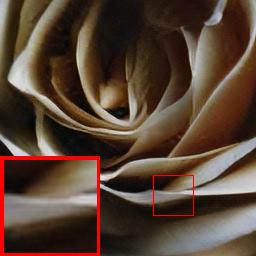} &
		\includegraphics[width=0.135\linewidth]{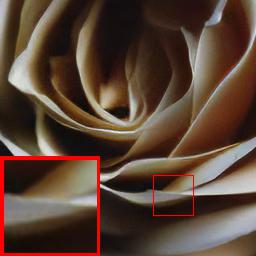} &
		\includegraphics[width=0.135\linewidth]{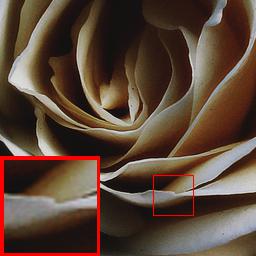} \\
		
		\small{ESRGAN-Bicubic} & \small{FSSR} & \small{Impressionism} & \small{DASR} & \small{DeFlow-NP} & \small{LUD-VAE} & \small{Ground Truth}
	\end{tabular}
	\caption{Visual comparison on real-world super-resolution dataset NTIRE20.}
	\label{rwsr_ntire}
\end{figure*}

\subsection{Results in real-world super-resolution}
We compare LUD-VAE with four unpaired degradation modeling methods namely FSSR~\cite{fritsche2019frequency} the winner of the AIM 2019 real-world super-resolution challenge~\cite{lugmayr2019aim}; Impressionism~\cite{ji2020real} the winner of the NTIRE 2020 real-world super-resolution challenge~\cite{lugmayr2020ntire}; DASR~\cite{wei2021unsupervised} a recently proposed GAN-based method; DeFlow~\cite{wolf2021deflow} a flow-based method. We use the settings without the pre-trained network to retrain the DeFlow model, denoted as DeFlow-NP. We use LUD-VAE and these four methods to learn the unknown degradation model, then downsample the high-resolution images and generate the low-resolution noisy images to obtain the paired training dataset. In addition, we use the downsampled data without degradation as the baseline, denoted as ESRGAN-Bicubic, see Figure~\ref{sr_gen}. When we have the training dataset, we use the real-world super-resolution model ESRGAN~\cite{wang2018esrgan} to obtain the final super-resolution results for all methods. We use the training code from Impressionism and train the ESRGAN model for 60k iterations and choose the final model with the best LPIPS score on the validation dataset after 50k, 60k iterations. We also compare LUD-VAE with the unsupervised super-resolution model CinCGAN~\cite{yuan2018unsupervised}, and two GAN-based models: Wang {\it et al.}~\cite{wang2021unsupervised}, and Yoon {\it et al.}~\cite{yoon2021simple}.

\begin{table*}
	\renewcommand{\arraystretch}{1.3}
	\setlength{\tabcolsep}{15pt}
	\centering
	\caption{Quantitative comparison on real-world image denoising dataset SIDD and DND. The synthetic datasets are trained with DnCNN~\cite{zhang2017beyond}.}
	\begin{tabular}{ccccccc}
		\hline
		\specialrule{0em}{.9pt}{.9pt}
		\rowcolor[gray]{0.95} 
		\cellcolor[gray]{0.95}                        & \multicolumn{2}{c}{\cellcolor[gray]{0.95}SIDD benchmark}        & \multicolumn{2}{c}{\cellcolor[gray]{0.95}SIDD validation} & \multicolumn{2}{c}{\cellcolor[gray]{0.95}DND benchmark}      \\
		\rowcolor[gray]{0.95} 
		\multirow{-2}{*}{\cellcolor[gray]{0.95}Method} & PSNR $\uparrow$                 & SSIM $\uparrow$                 & PSNR $\uparrow$                 & SSIM $\uparrow$    & PSNR $\uparrow$                 & SSIM $\uparrow$              \\ 
		\specialrule{0em}{.9pt}{.9pt}
		\hline
		CBM3D~\cite{dabov2007color}        & 25.65 & 0.685 & 25.65 & 0.475 & 34.51 & 0.851  \\
		Self2Self~\cite{quan2020self2self} & 29.51 & 0.651 & 29.46 & 0.595 & 35.19 & 0.868  \\
		MCWNNM~\cite{xu2017multi}          & 33.37 & 0.875 & 33.40 & 0.815 & 37.38 & 0.929  \\
		NC~\cite{lebrun2015noise}          & 31.26 & 0.826 & 31.31 & 0.725 & 35.43 & 0.884  \\
		C2N~\cite{jang2021c2n}             & 33.95 & 0.878 & 34.12 & 0.818 & 36.08 & 0.903  \\
		AWGN                               & 32.12 & 0.868 & 32.06 & 0.809 & 35.16 & 0.899  \\
		DeFlow-NP~\cite{wolf2021deflow}    & 33.54 & 0.875 & 33.53 & 0.817 & 36.48 & 0.916  \\
		\textbf{LUD-VAE} (ours)            & \textbf{34.82} & \textbf{0.926} & \textbf{34.91} & \textbf{0.892} & \textbf{37.60} & \textbf{0.933} \\ \hline
		DeFlow~\cite{wolf2021deflow}       & 33.81 & 0.897 & 33.82 & 0.846 & 36.71 & 0.923  \\
		DnCNN~\cite{zhang2017beyond}       & 36.54 & 0.927 & 36.83 & 0.870 & 37.90 & 0.943  \\ 
		\hline
	\end{tabular}
	\label{sidd_dnd}
\end{table*}

\begin{table}
	\renewcommand{\arraystretch}{1.3}
	\setlength{\tabcolsep}{12pt}
	\centering
	\caption{Quantitative comparison on low-light image enhancement dataset LOL. The synthetic datasets are trained with LLFlow~\cite{wang2021low}.}
	\begin{tabular}{cccc}
		\hline
		\specialrule{0em}{.9pt}{.9pt}
		\rowcolor[gray]{0.95} 
		Method                       & PSNR $\uparrow$ & SSIM $\uparrow$ & LPIPS $\downarrow$ \\ \hline
		\specialrule{0em}{.9pt}{.9pt}
		LIME~\cite{guo2016lime}                   & 16.55 & 0.4264 & 0.410              \\
		Zero-DCE~\cite{guo2020zero}               & 14.86 & 0.5588 & 0.335              \\
		SIEN~\cite{zhang2020self}                 & 19.50 & 0.7043 & 0.294              \\
		RUAS~\cite{liu2021retinex}                & 16.40 & 0.4996 & 0.270              \\
		SCI~\cite{ma2022toward}                   & 14.78 & 0.5220 & 0.339              \\
		EnlightenGAN~\cite{jiang2021enlightengan} & 17.48 & 0.6507 & 0.322              \\
		HEP~\cite{zhang2021unsupervised}          & 20.23 & 0.7910 & 0.167              \\
		LLFlow-GC                                 & 22.87 & 0.6222 & 0.306              \\
		\textbf{LUD-VAE} (ours)                   & \textbf{24.92} & \textbf{0.8502} & \textbf{0.157} \\ \hline
		RetinexNet~\cite{wei2018deep}             & 17.61 & 0.6479 & 0.386              \\
		KinD++~\cite{zhang2021beyond}             & 21.80 & 0.8338 & 0.158              \\ 
		LLFlow~\cite{wang2021low}                 & 25.00 & 0.8714 & 0.117              \\ \hline
	\end{tabular}
	\label{ll_lol}
\end{table}

The quantitative results are shown in Table~\ref{aim_ntire}. For the evaluation metrics, PSNR and SSIM focus on the restoration of the overall content of the image, while LPIPS pays more attention to the image details, so these two types of metrics are mutually exclusive from each other. Many methods perform well on only one metric type, whereas LUD-VAE performs uniformly well on these three metrics. For the AIM19 dataset, the degradation mainly behaves as low-frequency and structured noise, while the NTIRE20 dataset is primarily high-frequency noise. We found that LUD-VAE can learn these different degradation models without paired data. It is worth noting that the performance of ESRGAN-Bicubic, which only uses bicubic downsampling without any other degradation, is not satisfactory (Figure~\ref{rwsr_aim} and Figure~\ref{rwsr_ntire}). Thus,  the results illustrate the necessity of considering the noisy model in real-world super-resolution tasks. 

{\noindent \bf Comparison with flow-based methods.} Our model achieves comparable results with the DeFlow model. On the AIM19 dataset, our model outperforms DeFlow on the PSNR and LPIPS metrics, and on the NTIRE20 dataset, our model outperforms the DeFlow model on the SSIM metric. It is worth noting that the DeFLow model requires the use of a pre-trained model, which is trained on paired low-resolution and high-resolution data, which provides strong prior knowledge. Our method performs better than the DeFlow-NP model that does not use the pre-trained model. At the same time, the parameters of the DeFlow model are much larger than our model, which will bring difficulties to practical use, see Table~\ref{aim_np}. From the visual results in Figure~\ref{rwsr_aim} and Figure~\ref{rwsr_ntire}, our result removes the noise and is smoother and clearer than the DeFlow-NP model.

{\noindent \bf Comparison with GAN-based methods.} From Table~\ref{aim_ntire}, we find the performance of FSSR~\cite{fritsche2019frequency}, CinCGAN~\cite{yuan2018unsupervised}, and Wang {\it et al.}~\cite{wang2021unsupervised} on AIM19 dataset is better than NTIRE20 dataset, while the results of LUD-VAE and DeFlow~\cite{wolf2021deflow} are more consistent and robust. One possible reason is that GAN-based methods often need to fine-tune the weight among different loss functions. The other reason may be that the cycle-consistency constraint is not strong enough since the forward imaging model is ill-conditioned and the reconstruction requires high accuracy. In the LUD-VAE model, we do not use any heuristic loss function, such as perceptual loss or GAN-style loss. Besides, in Figure~\ref{rwsr_aim} and Figure~\ref{rwsr_ntire}, the results of FSSR~\cite{fritsche2019frequency} and DASR~\cite{wei2021unsupervised} are not sharp enough and still retain some noise speckles, while our results are more clear. 

\begin{figure*}
	\centering
	\begin{tabular}{c@{\hspace{0.005\linewidth}}c@{\hspace{0.005\linewidth}}c@{\hspace{0.005\linewidth}}c@{\hspace{0.005\linewidth}}c@{\hspace{0.005\linewidth}}c@{\hspace{0.005\linewidth}}c}
		\includegraphics[width=0.135\linewidth]{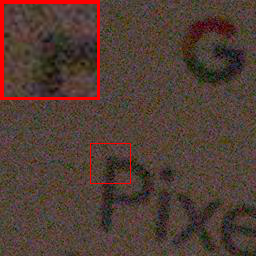}   &
		\includegraphics[width=0.135\linewidth]{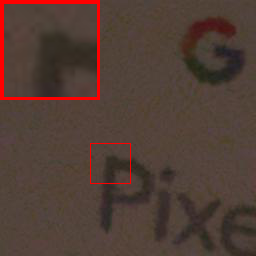}   &
		\includegraphics[width=0.135\linewidth]{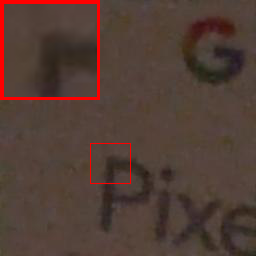} &
		\includegraphics[width=0.135\linewidth]{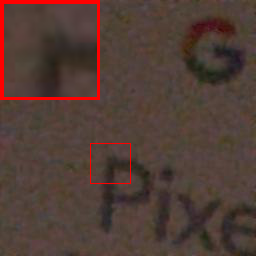} &
		\includegraphics[width=0.135\linewidth]{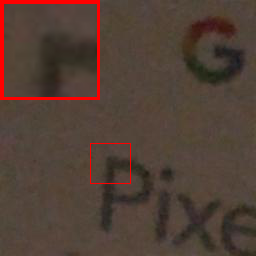} &
		\includegraphics[width=0.135\linewidth]{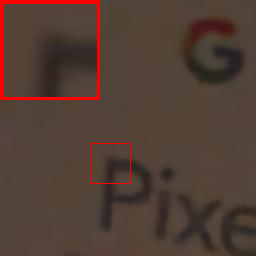} &
		\includegraphics[width=0.135\linewidth]{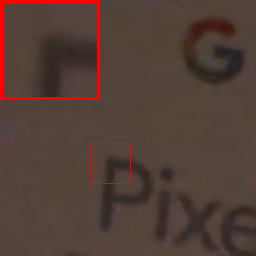} \\
		
		\includegraphics[width=0.135\linewidth]{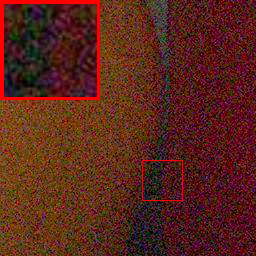}   &
		\includegraphics[width=0.135\linewidth]{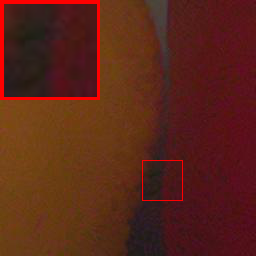}   &
		\includegraphics[width=0.135\linewidth]{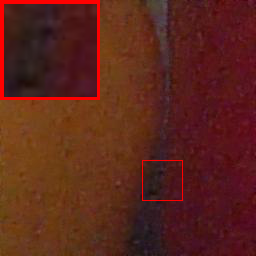} &
		\includegraphics[width=0.135\linewidth]{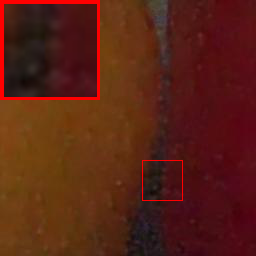} &
		\includegraphics[width=0.135\linewidth]{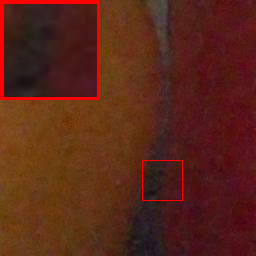} &
		\includegraphics[width=0.135\linewidth]{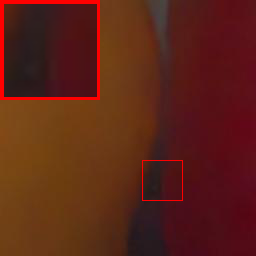} &
		\includegraphics[width=0.135\linewidth]{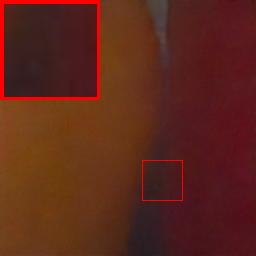} \\
		
		\includegraphics[width=0.135\linewidth]{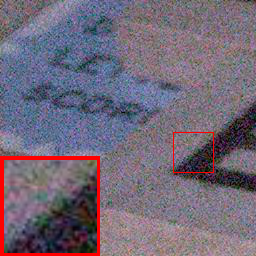}   &
		\includegraphics[width=0.135\linewidth]{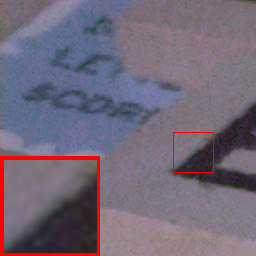}   &
		\includegraphics[width=0.135\linewidth]{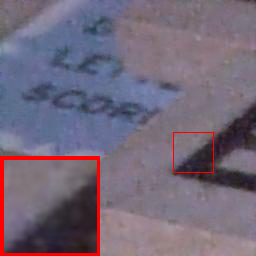} &
		\includegraphics[width=0.135\linewidth]{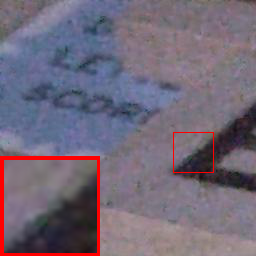} &
		\includegraphics[width=0.135\linewidth]{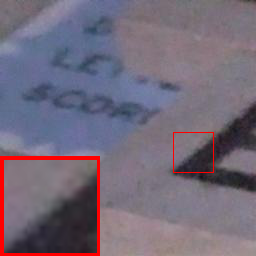} &
		\includegraphics[width=0.135\linewidth]{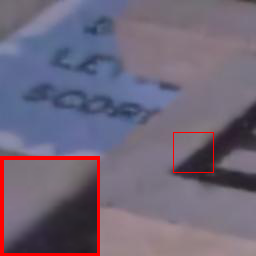} &
		\includegraphics[width=0.135\linewidth]{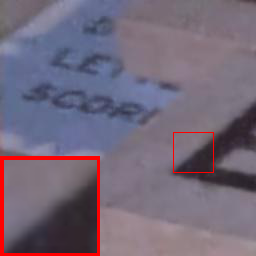} \\		
		
		\includegraphics[width=0.135\linewidth]{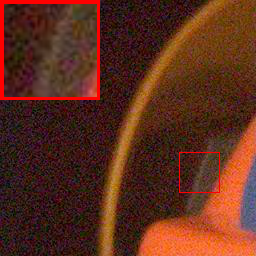}   &
		\includegraphics[width=0.135\linewidth]{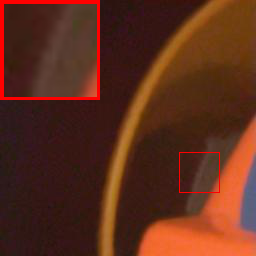}   &
		\includegraphics[width=0.135\linewidth]{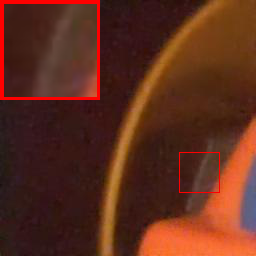} &
		\includegraphics[width=0.135\linewidth]{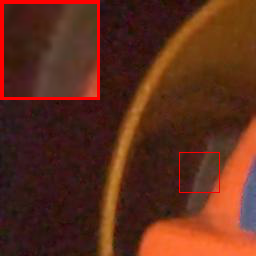} &
		\includegraphics[width=0.135\linewidth]{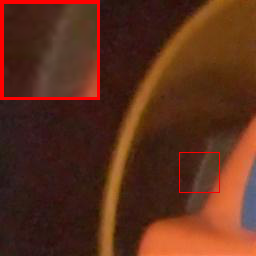} &
		\includegraphics[width=0.135\linewidth]{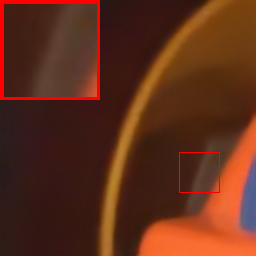} &
		\includegraphics[width=0.135\linewidth]{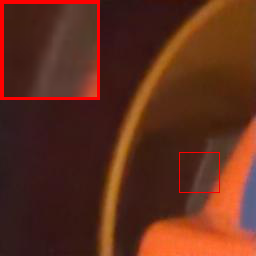} \\
		
		\includegraphics[width=0.135\linewidth]{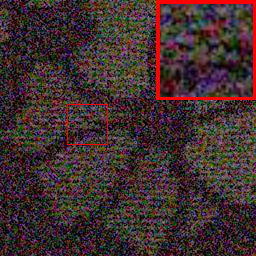} &
		\includegraphics[width=0.135\linewidth]{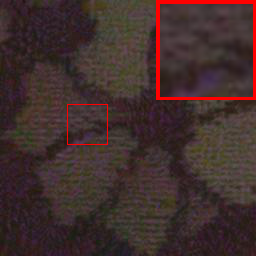} &
		\includegraphics[width=0.135\linewidth]{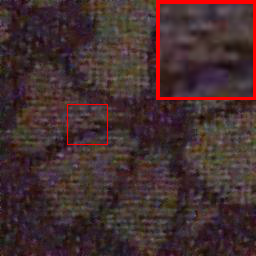} &
		\includegraphics[width=0.135\linewidth]{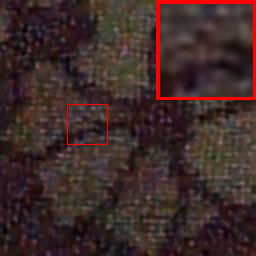} &
		\includegraphics[width=0.135\linewidth]{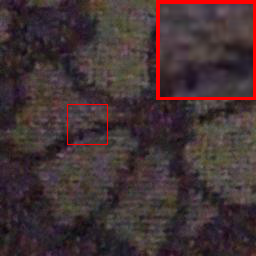} &
		\includegraphics[width=0.135\linewidth]{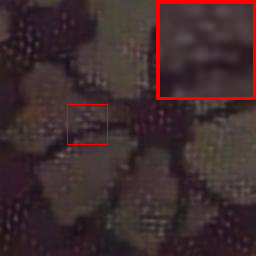} &
		\includegraphics[width=0.135\linewidth]{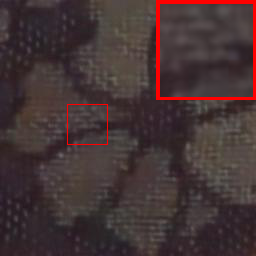} \\
		
		\small{Noisy} & \small{MCWNNM} & \small{C2N} & \small{AWGN} & \small{DeFlow-NP} & \small{LUD-VAE} & \small{DnCNN} \\
	\end{tabular}
	\caption{Visual comparison on real-world image denoising dataset SIDD benchmark.}
	\label{rwid_sidd}
\end{figure*}

\begin{figure*}
	\centering
	\begin{tabular}{c@{\hspace{0.01\linewidth}}c@{\hspace{0.01\linewidth}}c@{\hspace{0.01\linewidth}}c}
		\includegraphics[width=0.22\linewidth]{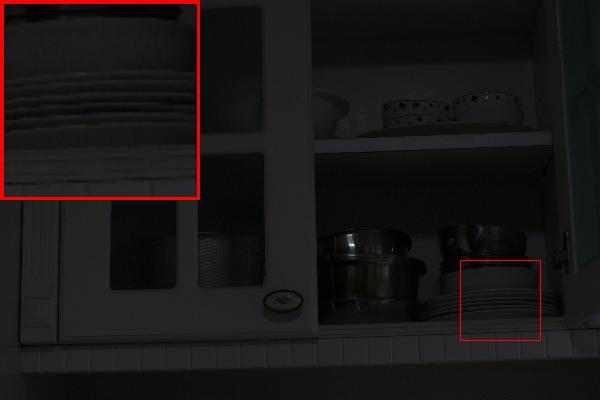} &
		\includegraphics[width=0.22\linewidth]{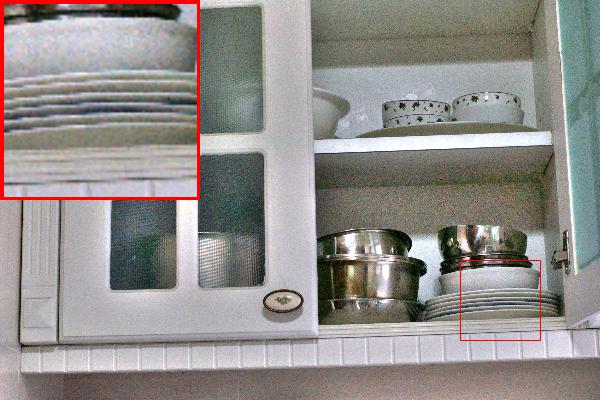} &
		\includegraphics[width=0.22\linewidth]{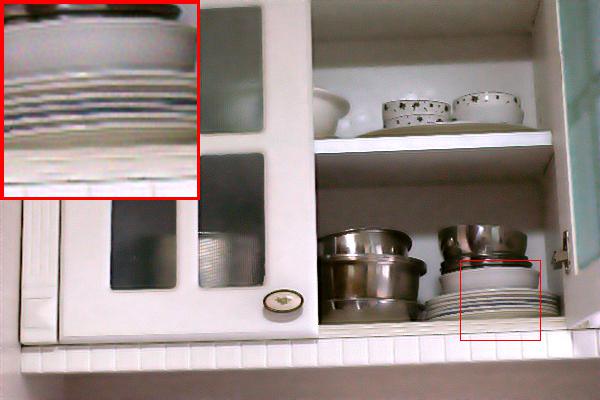} &
		\includegraphics[width=0.22\linewidth]{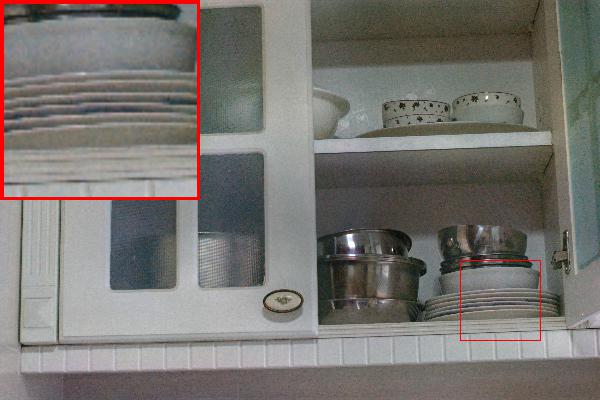} \\
		\small{Input} & \small{LIME} & \small{RUAS} & \small{EnlightenGAN} \\
		\includegraphics[width=0.22\linewidth]{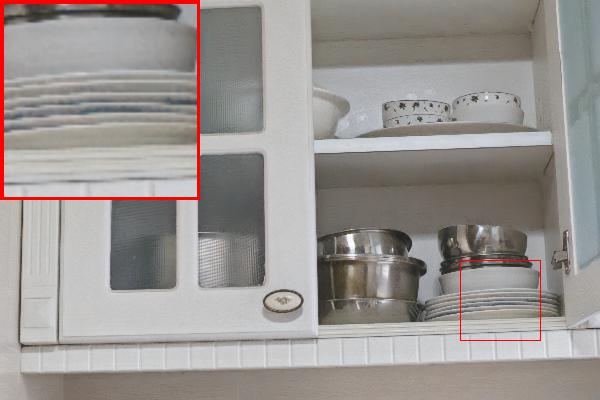} &
		\includegraphics[width=0.22\linewidth]{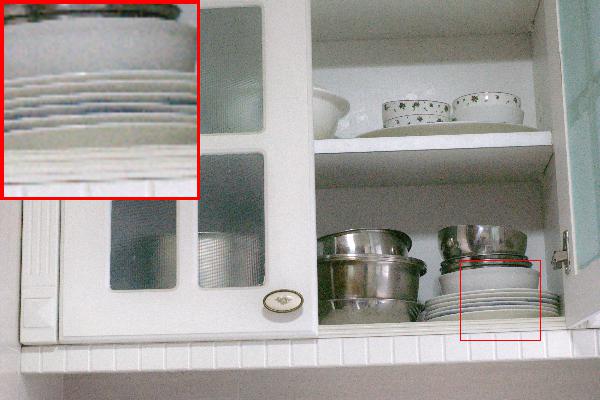} &
		\includegraphics[width=0.22\linewidth]{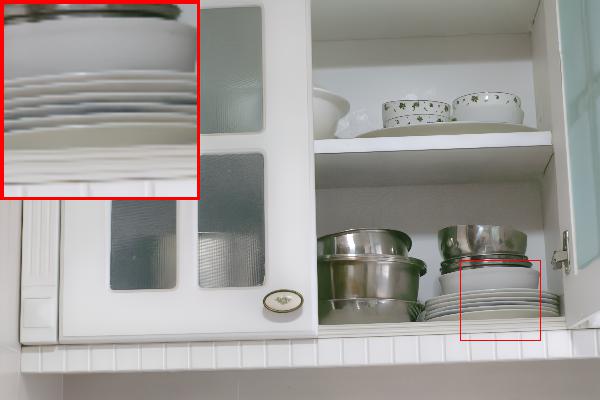} &
		\includegraphics[width=0.22\linewidth]{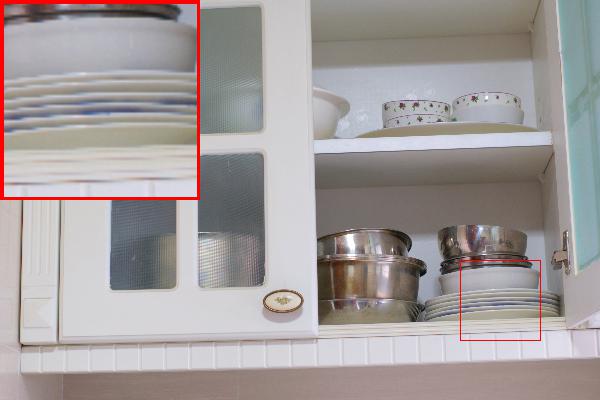} \\
		\small{HEP} & \small{LLFlow-GC} & \small{LUD-VAE} & \small{Ground Truth} \\
		
		\includegraphics[width=0.22\linewidth]{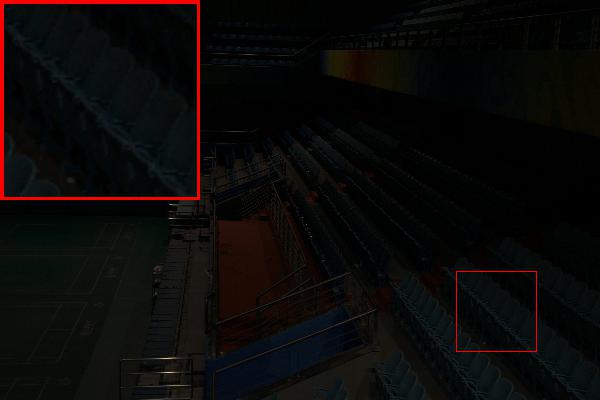} &
		\includegraphics[width=0.22\linewidth]{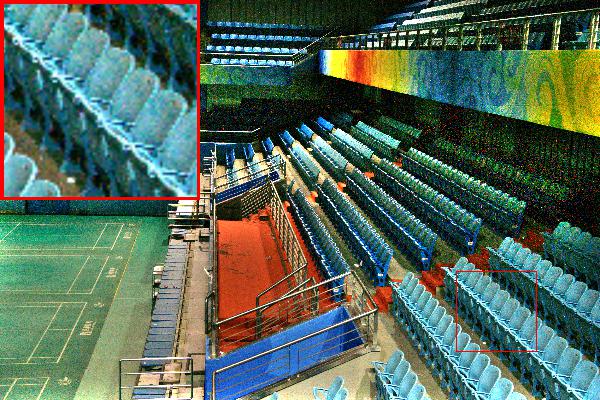} &
		\includegraphics[width=0.22\linewidth]{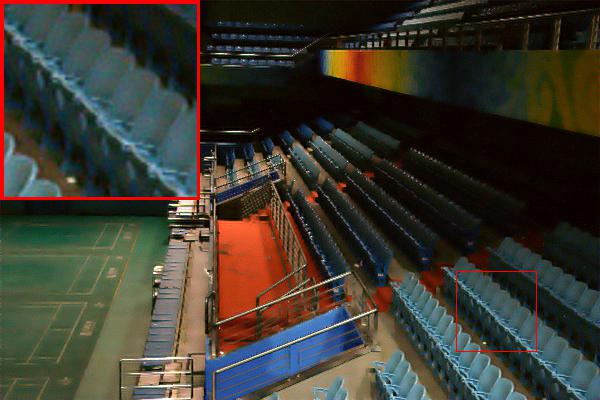} &
		\includegraphics[width=0.22\linewidth]{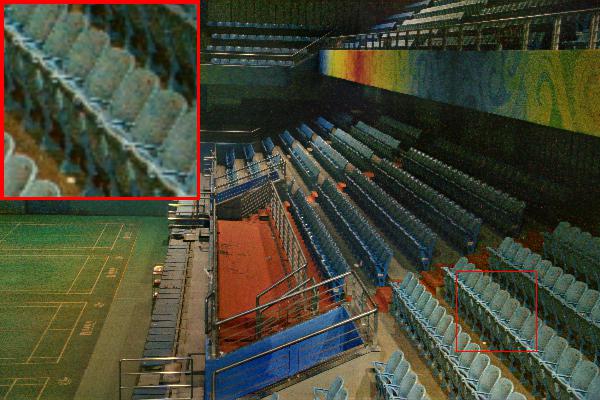} \\
		\small{Input} & \small{LIME} & \small{RUAS} & \small{EnlightenGAN} \\
		\includegraphics[width=0.22\linewidth]{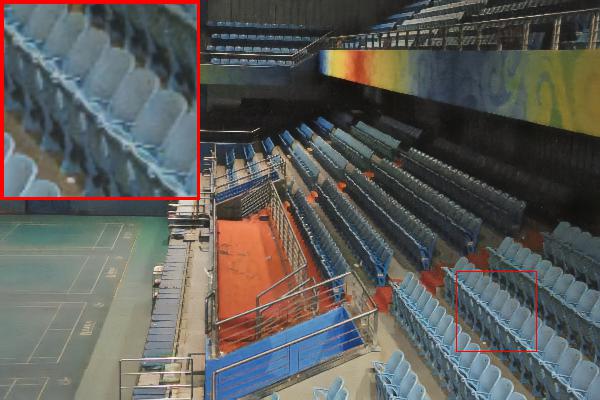} &
		\includegraphics[width=0.22\linewidth]{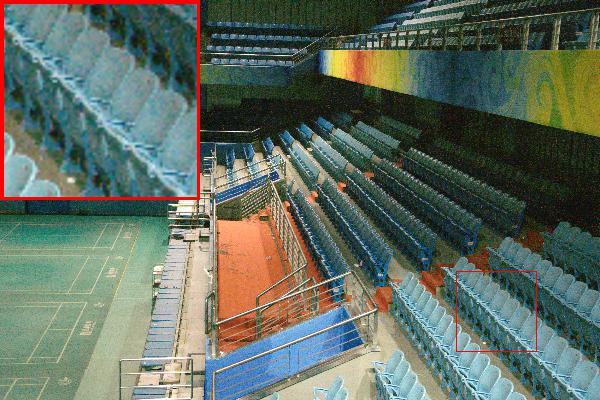} &
		\includegraphics[width=0.22\linewidth]{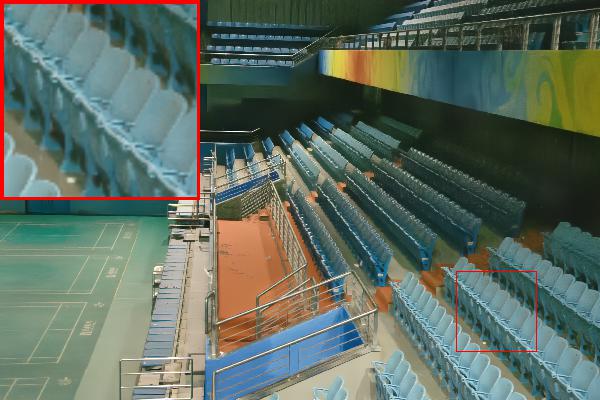} &
		\includegraphics[width=0.22\linewidth]{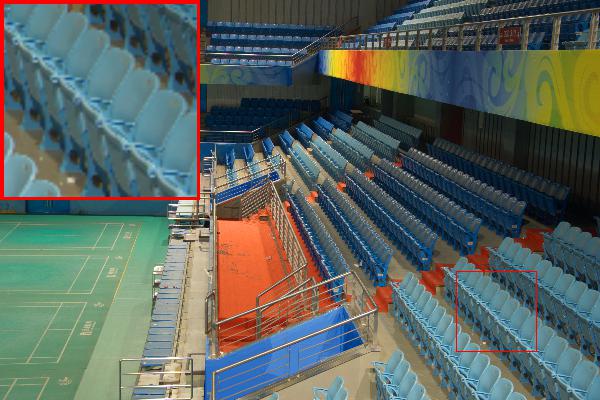} \\
		\small{HEP} & \small{LLFlow-GC} & \small{LUD-VAE} & \small{Ground Truth} \\

	\end{tabular}
	\caption{Visual comparison on low-light image enhancement dataset LOL.}
	\label{ll_lol_visual}
\end{figure*}

\subsection{Results in real-world image denoising}

We compare LUD-VAE with the unsupervised denoising method CBM3D~\cite{dabov2007color}, MCWNNM~\cite{xu2017multi}, NC~\cite{lebrun2015noise}, Self2Self~\cite{quan2020self2self}, the degradation modeling method DeFlow/DeFlow-NP~\cite{wolf2021deflow} and C2N~\cite{jang2021c2n}, and the fully supervised method DnCNN~\cite{zhang2017beyond}. We also set up a baseline method as additive white Gaussian noise degradation, denoted as AWGN. Since the SIDD dataset contains images with different noise levels, we synthetic the degraded images with different noise levels using different degradation methods. For AWGN, we randomly apply Gaussian noise with zero mean and standard deviation $\sigma \in [20, 120]$ to each image; for DeFLow/DeFlow-NP, we randomly apply synthetic noise with the noise level parameter $t \in [1,4]$ to each image; for LUD-VAE we choose $\sigma_\rvx \in [30,70]$ in~\eqref{iic} for each image randomly; For AWGN, LUD-VAE, DeFlow/DeFlow-NP, and C2N, we use the DnCNN~\cite{zhang2017beyond} for the downstream denoising tasks for 100k iterations with an initial learning rate $1e-4$ and halved in 50k iteration. We test the denoiser on the validation set every 500 iterations, and choose the model with the best PSNR to evaluate on the benchmark set.

The quantitative results are shown in Table~\ref{sidd_dnd}. We find that LUD-VAE has achieved the best results except for the supervised method DnCNN. DeFlow does not perform very well on this dataset; the reason may be the incomplete loss function of the model (see Appendix) and the instability of the training process. The visual results in Figure~\ref{rwid_sidd} show that our results are closer to DnCNN's, with complete noise removal compared with the other methods.

\subsection{Results in low-light image enhancement}

We compare LUD-VAE with unsupervised low-light image enhancement method LIME~\cite{guo2016lime}, Zero-DCE~\cite{guo2020zero}, RUAS~\cite{liu2021retinex}, SCI~\cite{ma2022toward}, SIEN~\cite{zhang2020self}, and unpaired modeling method HEP~\cite{zhang2021unsupervised}, EnlightenGAN~\cite{jiang2021enlightengan}, and supervised method RetinexNet~\cite{wei2018deep}, KinD++~\cite{zhang2021beyond}, and LLFlow~\cite{wang2021low}. We set up a baseline method with only gamma correction as the degradation model, denoted as LLFlow-GC. For LUD-VAE and LLFlow-GC, we use the supervised method LLFlow~\cite{wang2021low} for the downstream restoration model. All compared methods are evaluated with their official codes.

The quantitative results are shown in Table~\ref{ll_lol}. The LUD-VAE achieves the best results compared with all unsupervised and unpaired methods on all three quantitative matrices. The performance of the gamma correction model LLFlow-GC declined significantly compared with LUD-VAE, indicating that the simulated degradation of LUD-VAE is more accurate. Visual results are shown in Figure~\ref{ll_lol_visual}, where we find our method removes the noise more completely compared with the other methods.

\subsection{Ablation study and discussions}

{\noindent \bf Generation method.} We compare different methods in Figure~\ref{three_methods} to generate paired data in our model. For L2CN, we need to train an additional PixelCNN~\cite{oord2016conditional} to sample from the prior distribution $p(\rvz)$, which is inconvenient for practical use; thus, we do not adopt this method. For N2C and C2N, we evaluate them on the SIDD dataset. The results are shown in Table~\ref{sidd_m2_m3}, and we find that C2N achieves better performance than N2C. Here, we give some explanations that possibly show that C2N is more reliable than N2C. It is noted that N2C uses the noisy image $\rvy$ to infer the latent variable $\rvz$, while C2N uses a clean image $\rvx$ to infer $\rvz$. By the generative graph, the clean image $\rvx$ only depends on $\rvz$ while the noisy image $\rvy$ depends on $\rvz$ and $\rvz_\rvn$. Thus, C2N may be more robust than N2C in inferring latent variable $\rvz$. Besides, C2N simulates the degradation process to the clean image, which is a forward process. On the other hand, N2C estimates the corresponding clean images, which is an inverse process. It is clear that simulating the inverse process is more difficult than the forward process and has the stability issue for the ill-posed problems. Thus, C2N is relatively easier to obtain reliable paired data. In addition, N2C is a deterministic mapping, while C2N is stochastic. Since one clean image has many different degraded counterparts, C2N is more suitable for the downstream denoising model.

\begin{table}
	\renewcommand{\arraystretch}{1.3}
	\setlength{\tabcolsep}{10pt}
	\centering
	\caption{Different generation methods on SIDD dataset.}
	\begin{tabular}{ccccc}
		\hline
		\specialrule{0em}{.9pt}{.9pt}
		\rowcolor[gray]{0.95} 
		\cellcolor[gray]{0.95}                         & \multicolumn{2}{c}{\cellcolor[gray]{0.95}SIDD benchmark} & \multicolumn{2}{c}{\cellcolor[gray]{0.95}SIDD validation} \\
		\rowcolor[gray]{0.95} 
		\multirow{-2}{*}{\cellcolor[gray]{0.95}Method} & PSNR $\uparrow$              & SSIM $\uparrow$             & PSNR $\uparrow$              & SSIM $\uparrow$              \\ 
		\specialrule{0em}{.9pt}{.9pt}
		\hline
		N2C                               & 34.74 & 0.901 & 34.78 & 0.858 \\
		\textbf{C2N}                      & 34.82 & 0.926 & 34.91 & 0.892 \\ \hline
	\end{tabular}
	\label{sidd_m2_m3}
\end{table}

\begin{table}
	\renewcommand{\arraystretch}{1.3}
	\setlength{\tabcolsep}{9pt}
	\centering
	\caption{Validate inference invariant condition on AIM19 dataset.}
	\begin{tabular}{cccccc}
		\hline
		\specialrule{0em}{.9pt}{.9pt}
		\rowcolor[gray]{0.95} 
		$\sigma_\rvx$                & $\sigma_\rvy$                & PSNR $\uparrow$ & SSIM $\uparrow$ & LPIPS $\downarrow$ & MMD \\ 
		\specialrule{0em}{.9pt}{.9pt}
		\hline
		0                            & 0                            & 21.66           & 0.5349          & 0.519     & 7.64         \\
		10                           & 5                            & 21.87           & 0.5510          & 0.374     & 4.42         \\
		\textbf{15}                  & \textbf{10}                  & 22.32           & 0.6197          & 0.341     & 2.34         \\
		20                           & 10                           & 22.47           & 0.6355          & 0.389     & 2.08         \\
		15                           & 5                            & 22.96           & 0.6492          & 0.402     & 4.66         \\
		20                           & 15                           & 22.20           & 0.6199          & 0.343     & 1.35         \\ \hline
	\end{tabular}
	\label{aim_iic}
\end{table}

\begin{table}
	\renewcommand{\arraystretch}{1.3}
	\setlength{\tabcolsep}{7pt}
	\centering
	\caption{Different stochastic layers of LUD-VAE on AIM19 dataset.}
	\begin{tabular}{cccccc}
		\hline
		\specialrule{0em}{.9pt}{.9pt}
		\rowcolor[gray]{0.95} 
		\cellcolor[gray]{0.95} & & & & \#Params & Times     \\
		\rowcolor[gray]{0.95} 
		\multirow{-2}{*}{\cellcolor[gray]{0.95}\#Layers} & \multirow{-2}{*}{\cellcolor[gray]{0.95}PSNR}& \multirow{-2}{*}{\cellcolor[gray]{0.95}SSIM} & \multirow{-2}{*}{\cellcolor[gray]{0.95}LPIPS}  & (Million)    & (Seconds) \\ 
		\specialrule{0em}{.9pt}{.9pt}
		\hline
		1 Layer            & 21.24 & 0.5364 & 0.375 & 1.83     & 0.347 \\
		2 Layers           & 21.96 & 0.6139 & 0.344 & 3.77     & 0.365 \\
		\textbf{3 Layers}  & 22.32 & 0.6197 & 0.341 & 5.71     & 0.387 \\
		4 Layers           & 22.14 & 0.6209 & 0.346 & 7.65     & 0.394 \\
		5 Layers           & 22.07 & 0.6137 & 0.341 & 9.58     & 0.409 \\ \hline
	\end{tabular}
	\label{aim_nl}
\end{table}

\begin{table}
	\renewcommand{\arraystretch}{1.3}
	\setlength{\tabcolsep}{16pt}
	\centering
	\caption{Model parameters and GPU running time.}
	\begin{tabular}{ccc}
		\hline
		\specialrule{0em}{.9pt}{.9pt}
		\rowcolor[gray]{0.95} 
		\cellcolor[gray]{0.95}                          & \#Params      & Times     \\
		\rowcolor[gray]{0.95} 
		\multirow{-2}{*}{\cellcolor[gray]{0.95}Method}   & (Million)    & (Seconds) \\ 
		\specialrule{0em}{.9pt}{.9pt}
		\hline
		FSSR~\cite{fritsche2019frequency}          & 1.62         & 0.011     \\
		DASR~\cite{wei2021unsupervised}                 & 1.70         & 0.011     \\
		DeFlow~\cite{wolf2021deflow}                    & 62.94        & 2.548     \\
		\textbf{LUD-VAE} (ours)                         & 5.71         & 0.387     \\ \hline
	\end{tabular}
	\label{aim_np}
\end{table}

\begin{table}
	\renewcommand{\arraystretch}{1.3}
	\setlength{\tabcolsep}{9pt}
	\centering
	\caption{Different downsample methods on AIM19 dataset.}
	\begin{tabular}{cccc}
		\hline
		\specialrule{0em}{.9pt}{.9pt}
		\rowcolor[gray]{0.95} 
		Downsample method                & PSNR $\uparrow$ & SSIM $\uparrow$ & LPIPS $\downarrow$ \\ 
		\specialrule{0em}{.9pt}{.9pt}
		\hline
		Nearest                        & 22.65           & 0.6319          & 0.351              \\
		Bilinear                       & 21.62           & 0.6137          & 0.338              \\
		Bicubic                        & 22.32           & 0.6197          & 0.341              \\ \hline
	\end{tabular}
	\label{downsample_method}
\end{table}

{\noindent \bf Inference invariant condition.} We verify the inference invariant condition on the AIM19 dataset. Firstly, we evaluate our method with different pre-processing noise levels $\sigma_\rvx$ and $\sigma_\rvy$, and then use the Maximum Mean Discrepancy~\cite{gretton2012kernel} (MMD) to measure the distance between $q(\rvz|\rvx)$ and $q(\rvz|\rvy)$ for validating the establishment of the Inference invariant condition~\eqref{inferece_assume}. We use the validation set of AIM19 and central crop the images to size $128\times128$. For each image, we draw 10 samples $\{\rvz^i|\rvx\}_{i=1}^{10}$, $\{\rvz^i|\rvy\}_{i=1}^{10}$ from $q(\rvz|\rvx)$, $q(\rvz|\rvy)$ respectively, and then calculate MMD metric between $\{\rvz^i|\rvx\}_{i=1}^{10}$ and $\{\rvz^i|\rvy\}_{i=1}^{10}$. We compute the average MMD metric on the whole validation set, and the results are shown in Table~\ref{aim_iic}. From the table, we find that in case $\sigma_\rvx=0$, $\sigma_\rvy=0$, the MMD is the largest, which means the inference invariant condition does not hold, and the model cannot learn the unknown degradation process, resulting in poor performance. In addition, the performance of $\sigma_\rvy=10$, $15$ is better than $\sigma_\rvy=5$. One possible explanation is that when $\sigma_\rvy$ is relatively large, the Gaussian noise will overwhelm the original unknown noise, making the inference invariant condition more satisfied, which is shown as the lower MMD metrics. Meanwhile, comparing the results of $\sigma_\rvx=15$, $\sigma_\rvy=10$ with $\sigma_\rvx=20$, $\sigma_\rvy=10$, we find that increasing the gap between $\sigma_\rvx$ and $\sigma_\rvy$, which means the degradation level of $\rvy$ is larger will make LPIPS worse and make PSNR and SSIM better. This may be because when the degradation level of $\rvy$ is chosen to exceed the real degradation level, our model will synthesize higher degraded images than the real degraded images. The denoising performance of the image restoration model will trend to make the image smoother and lose the details. Increasing the pre-processing noise level can decrease the MMD metric, making the inference invariant condition more established. However, injecting higher-level noise will impair more image information, reducing the method's representation ability. Therefore, we need to balance inference invariant condition with an appropriate level of pre-processing noise and the model representation ability in practical applications.

{\noindent \bf Stochastic latent layer.} We investigate the structure of our hierarchical VAE model with different stochastic layers, see Table~\ref{aim_nl} for the results. From the table, we find that the hierarchical VAE model outperforms the single layer structure significantly, indicating the necessity of using the hierarchical structure. Moreover, there is little difference in the final performance when the latent depth is larger than three. To reduce the number of parameters, our model adopts the three-layer structure. 

{\noindent \bf Degradation operation.} Since we assumes that the approximated degradation model is known in image restoration tasks, we test the sensitivity of the proposed method to different degradation strategies. In particular, we test three different downsample method: nearest, bilinear and bicubic as the degradation mode in the image super-resolution task. We implement downsample operations with matlab imresize function. The results are shown in Table~\ref{downsample_method}. From the table, we find the nearest downsample has the best performance on PSNR and SSIM metrics among these three methods, while the bilinear downsample has the best perceptual score. From the perspective of the perceptual score LPIPS, the difference between these three methods is lower than 0.015, which shows our method is robust to different downsample operators.

{\noindent \bf Model parameter.} We compare the model parameters and the GPU running time for generating paired data (the dimension of the input image is $256 \times 256 \times 3$) of different unpaired learning models, see Table~\ref{aim_np} for the results. The table shows that DASR and FSSR have the lowest model parameters and running times, but their restoration performance is beneath DeFlow and LUD-VAE methods. Compared with DeFlow, our model has much fewer parameters and running time, which is convenient for practical uses. 

\section{Conclusion and future work}
This paper proposes LUD-VAE, a degradation modeling method using unpaired data based on variational inference. We disentangle the clean and corrupted data domains through a probabilistic graphical model, which enables the transformation from clean to corrupted data by estimating the joint probability density function of the two domains using unpaired data from each domain individually. Moreover, we establish the equivalency between paired and unpaired learning for LUD-VAE under the inference invariant condition, which provides the mathematical rationale for our approach. We use LUD-VAE to generate synthetic training datasets for downstream supervised learning methods and evaluate them on real-world denoising, super-resolution, and low-light image enhancement tasks. Experimental results show that our method achieves state-of-the-art results on real-world image datasets.

In this paper, we assume that the approximated degradation process is known, which can be derived from the physical properties of the imaging process. However, the degradation process is unavailable in image-to-image translation problems such as style transfer; thus, we can not directly apply our model to such problems. Besides, similar issues exist for the multi-modal tasks, including text and image translation. Therefore, one possible solution is to use the GAN model to simulate the degradation process, develop the registration methods among two domains, and then apply the LUD-VAE method. Thus, extending LUD-VAE to border applications is our future research direction.

\section*{Acknowledgments}
This work was supported by the National Key R\&D Program of China (No. 2021YFA1001300), National Natural Science Foundation of China (No.11901338), Tsinghua University Initiative Scientific Research Program.

\appendices

\section{Discussions on maximum likelihhood}

\begin{figure}[H]
	\centering
	\includegraphics[width=0.33\linewidth]{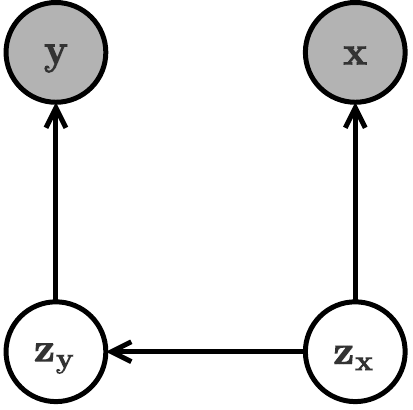}
	\caption{Generative process in DeFlow model.}
	\label{graph_model_deflow}
\end{figure}

Considering the generative model in Figure~\ref{graph_model_deflow} that is suggested by the DeFlow method, where $\rvz_\rvx\sim \gN(0, I)$, $\rvz_\rvy=\rvz_\rvx+\rvu$, $\rvu\sim \gN(\mu_{u}, \Sigma_{u})$. 
In the DeFlow method, it maximizes the log-likelihood function of the two marginal densities:
\begin{equation}\label{DeFlow}
	\max_{\theta} \sum_{i} \log p(\rvx_{i}) + \sum_{j} \log p(\rvy_{j}).
\end{equation}

where
\begin{equation}
	\begin{aligned}
		&\log p(\rvx)=\log \left|\operatorname{det} D f_{\theta}(\rvx)\right| + \log \mathcal{N}\left(f_{\theta}(\rvx) ; 0, I\right) \\
		&\log p(\rvy)=\log \left|\operatorname{det} D f_{\theta}(\rvy)\right| + \log \mathcal{N}\left(f_{\theta}(\rvy) ; \mu_{u}, I+\Sigma_{u}\right)
	\end{aligned}
\end{equation}
by the change of variables formula, where $f_{\theta}$ is an invertible normalizing flow. In the next, we show that the objective function in~\eqref{DeFlow} is incomplete for representing the log-likelihood $\log p(\rvx,\rvy)$. 
Define
\begin{equation}
	F(\rvx, \rvy) = (f_{\theta}(\rvx), f_{\theta}(\rvy)) = (\rvz_\rvx,\rvz_\rvy),
\end{equation}
where $(\rvz_\rvx,\rvz_\rvy)\sim \gN\left(\left[\begin{matrix} 0 \\ \mu_{u} \end{matrix} \right], \left[\begin{matrix}I & I \\ I & I +\Sigma_{u}\end{matrix}\right] \right)$ as $\rvz_\rvy = \rvz_\rvx + \rvu$. Using the change of variables formula, we have
\begin{equation}
	p(\rvx, \rvy) = \left|\operatorname{det} D F(\rvx, \rvy)\right| \gN\left(F(\rvx, \rvy) ; \left[\begin{matrix} 0 \\ \mu_{u} \end{matrix} \right], \left[\begin{matrix}I & I \\ I & I +\Sigma_{u}\end{matrix}\right] \right).
\end{equation}
Since 
\begin{equation}
	F \left( \left[\begin{matrix}\rvx \\ \rvy\end{matrix}\right] \right)= \left[\begin{matrix} f_{\theta}(\rvx) \\ f_{\theta}(\rvy)\end{matrix}\right] \Rightarrow D F\left(\left[\begin{matrix}\rvx \\ \rvy\end{matrix}\right]\right) = \left[\begin{matrix}Df_{\theta}(\rvx) & 0 \\ 0 & Df_{\theta}(\rvy)\end{matrix}\right],
\end{equation}
we have
\begin{equation}
	\begin{aligned}
		&p(\rvx, \rvy) = \frac{\left|\operatorname{det} D f_{\theta}(\rvx)\right|\left|\operatorname{det} D f_{\theta}(\rvy)\right|}{\sqrt{(2\pi)^{2n} \det \Sigma_{u}}} \times \\
		&\exp \left\{-\frac{1}{2} \left[\begin{matrix}f_{\theta}(\rvx) \\ f_{\theta}(\rvy) - \mu_{u}\end{matrix}\right]^{T} \left[\begin{matrix}I+\Sigma_{u}^{-1} & -\Sigma_{u}^{-1} \\ -\Sigma_{u}^{-1} & \Sigma_{u}^{-1}\end{matrix}\right] \left[\begin{matrix}f_{\theta}(\rvx) \\ f_{\theta}(\rvy) - \mu_{u}\end{matrix}\right] \right\},
	\end{aligned}
\end{equation}
where $n$ is the dimension of random variables $\rvz_\rvx$ and $\rvz_\rvy$. Then the log-likelihood function $\log p(\rvx, \rvy)$ can be decomposed into
\begin{equation}
	\begin{aligned}
		\log p(\rvx, \rvy) =& \log \mathcal{N}\left(f_{\theta}(\rvx) ; 0, I\right) + \log \left|\operatorname{det} D f_{\theta}(\rvx)\right| \\
		&+ \log \mathcal{N}\left(f_{\theta}(\rvy) ; \mu_{u}, I+\Sigma_{u}\right) + \log \left|\operatorname{det} D f_{\theta}(\rvy)\right| \\
		&+ \log \gN \left(f_{\theta}(\rvx); -\mu_{u},\Sigma_{u}\right) + f_{\theta}^{T}(\rvx) \Sigma_{u}^{-1} f_{\theta}(\rvy) \\
		&+ \frac{1}{2}\log \left((2\pi)^{n} \operatorname{det} (\Sigma_{u})\right) - \frac{1}{2} \mu_{u}^{T} \Sigma_{u}^{-1} \mu_{u}. 
	\end{aligned}
\end{equation}
Using marginal distribution notation, the maximal likelihood estimation is
\begin{equation}\label{deflow_likelihood}
	\begin{aligned}
		&\max_{\theta} \sum_{i} \log p(\rvx_{i}, \rvy_{i}) \\
		=& \max_{\theta} \sum_{i} \log p(\rvx_{i}) + \sum_{j} \log p(\rvy_{j}) \\
		&+\sum_{i} \log \gN \left(f_{\theta}(\rvx_{i}); -\mu_{u},\Sigma_{u}\right) + f_{\theta}^{T}(\rvx_{i}) \Sigma_{u}^{-1} f_{\theta}(\rvy_{i}) \\
		&+ \frac{1}{2}\log \left((2\pi)^{n} \operatorname{det} (\Sigma_{u})\right) - \frac{1}{2} \mu_{u}^{T} \Sigma_{u}^{-1} \mu_{u}. 
	\end{aligned}
\end{equation}
The additional term $\sum_{i} f_{\theta}^{T}(\rvx_{i}) \Sigma_{u}^{-1} f_{\theta}(\rvy_{i})$ in~\eqref{deflow_likelihood} requires the paired information. But, in the DeFlow model, it further introduces conditional marginal likelihood and its relationship with conditional likelihood is still unknown and deserves further exploration. Inspired by the above derivations, we propose our generative graph in which the two latent variables are independent that is relatively easy for constructing an approximation of the log-likelihood.

\section{Mathematical details on inference invariant condition}

We provide more detailed derivations on the inference invariant condition.

\subsection{The design of pre-process operator}
The pre-process operator $h$ is defined as adding Gaussian noise to the input data, {\it i.e.,}
\begin{equation}\label{q41}
	h(\rvx) = \rvx + \rvn_\rvx, \quad h(\rvy) = \rvy + \rvn_\rvy,
\end{equation}
where $\rvn_\rvx \sim \gN(\mathbf{0}, \sigma_\rvx^2 \rmI)$ and $\rvn_\rvy \sim \gN(\mathbf{0}, \sigma_\rvy^2 \rmI)$. So the encoding procedure $q(\rvz | \rvx)$ consists of two steps
\begin{itemize}
	\item[(1)] Sample $\rvn_\rvx \sim \gN(\mathbf{0}, \sigma_\rvx^2 \rmI)$.
	\item[(2)] Feed $\rvx + \rvn_\rvx$ into the encoding network.
\end{itemize}
The encoding procedure for $q(\rvz | \rvy)$ is the same as $q(\rvz|\rvx)$.

\subsection{The derivation of $q(\rvz | \rvx)$ and $q(\rvz | \rvy)$}
First we have
\begin{equation}\label{q42}
	q(\rvz|\rvx) = \int q(\rvz, \rvn_\rvx | \rvx) d\rvn_\rvx = \int q(\rvn_\rvx | \rvx) q(\rvz | \rvx, \rvn_\rvx)d\rvn_\rvx,
\end{equation}
and since the Gaussian noise $\rvn_\rvx$ is independent from the input data $\rvx$, so $q(\rvn_\rvx | \rvx) = p(\rvn_\rvx)$, and 
\begin{equation}\label{q43}
	\begin{aligned}
		\int q(\rvn_\rvx | \rvx) q(\rvz | \rvx, \rvn_\rvx)d\rvn_\rvx &= \int p(\rvn_\rvx) q(\rvz | \rvx, \rvn_\rvx) d\rvn_\rvx \\
		& = \E_{p(\rvn_\rvx)} q(\rvz | \rvx, \rvn_\rvx).
	\end{aligned}
\end{equation}
Choosing $q(\rvz | \rvx, \rvn_\rvx) = q(\rvz | \rvx + \rvn_\rvx)$ and $q(\rvz | \rvy, \rvn_\rvy) = q(\rvz | \rvy + \rvn_\rvy)$, it has
\begin{equation}\label{q45}
	q(\rvz | \rvx) = \E_{p(\rvn_\rvx)} q(\rvz | \rvx + \rvn_\rvx), \quad q(\rvz | \rvy) = \E_{p(\rvn_\rvy)} q(\rvz | \rvy + \rvn_\rvy).
\end{equation}
One typical choice of $q(\rvz|\rvx,\rvn_\rvx)$ is $\delta(e^l(\rvx + \rvn_\rvx))$ where $e^l$ is the encoding networks and $\delta(\cdot)$ denotes the delta distribution. Together with the assumption that $\rvy = \rvx + \rvn$, we have

\begin{equation}\label{q46}
	\E_{p(\rvn_\rvy)} q(\rvz | \rvy + \rvn_\rvy) = \E_{p(\rvn_\rvy)} q(\rvz | \rvx + \rvn + \rvn_\rvy) = \E_{p(\tilde{\rvn}_\rvy)} q(\rvz | \rvx + \tilde{\rvn}_\rvy),
\end{equation}
where $\tilde{\rvn}_\rvy = \rvn + \rvn_\rvy \sim \gN(\rvn, \sigma_\rvy^2 \rmI)$. Combining \eqref{q45} and \eqref{q46}, we have
\begin{equation}
	q(\rvz | \rvy) = \E_{p(\tilde{\rvn}_\rvy)} q(\rvz | \rvx + \tilde{\rvn}_\rvy).
\end{equation}

\subsection{The difference between $q(\rvz|\rvx)$ and $q(\rvz|\rvy)$}
From the above derivation, for paired $\rvx$ and $\rvy$, we have
\begin{equation}
	q(\rvz | \rvx) = \E_{p(\rvn_\rvx)} q(\rvz | \rvx + \rvn_\rvx), \quad q(\rvz | \rvy) = \E_{p(\tilde{\rvn}_\rvy)} q(\rvz | \rvx + \tilde{\rvn}_\rvy).
\end{equation}
Assume $p_1(\rvn)$, $p_2(\rvn)$ to be the density function of $\tilde{\rvn}_\rvy$ and $\rvn_\rvx$, respectively, then
\begin{equation}\label{q47}
	\begin{aligned}
		\| q(\rvz | \rvx) - q(\rvz | \rvy) \|_1 &= \int \left| \int (p_1(\rvn) - p_2(\rvn) )q(\rvz | \rvx + \rvn) d\rvn \right| d\rvz \\
		&\leq \int \int \left| (p_1(\rvn) - p_2(\rvn) ) \right| q(\rvz | \rvx + \rvn) d\rvn  d\rvz \\
		&= \int \left| (p_1(\rvn) - p_2(\rvn) ) \right| \int q(\rvz | \rvx + \rvn) d\rvz d\rvn \\
		&= \int \left| (p_1(\rvn) - p_2(\rvn) ) \right| d\rvn \\
		&= \| p_1(\rvn) - p_2(\rvn) \|_1.
	\end{aligned}
\end{equation}
From~\cite{cover1999elements} (Lemma 12.6.1) we have:
\begin{equation}\label{q48}
	\begin{aligned}
		\| p_1(\rvn) - p_2(\rvn) \|_1 &\leq \sqrt{2\log2 \KL(p_1 \| p_2)} \\
		&= \sqrt{2\log2 \KL(p(\tilde{\rvn}_\rvy) \| p(\rvn_\rvx) )},
	\end{aligned}
\end{equation}
and
\begin{equation}\label{q49}
	\KL(p(\tilde{\rvn}_\rvy) \| p(\rvn_\rvx) )  = \log \frac{ \sigma^{K}_{\rvx} }{ \sigma^{K}_{\rvy} } - \frac{K}{2} +  \frac{K\sigma_\rvy^2}{2\sigma_\rvx^2} + \frac{\|\rvn \|_2^2}{2\sigma_\rvx^2} \to 0,
\end{equation}
as $\sigma_\rvx, \sigma_\rvy \to \infty$, where $K$ is the dimension of random variables.
Combining \eqref{q47}, \eqref{q48}, and \eqref{q49}, we have $\| q(\rvz | \rvx) - q(\rvz | \rvy) \|_1$ approaches to $0$ as $\sigma_\rvx,\sigma_\rvy \to \infty$ and $\lim\limits_{\sigma_\rvx,\sigma_\rvy\to \infty}\sigma_\rvx/\sigma_\rvy = 1$. In this case, we know the inference invariant condition approximately holds.

%
\bibliographystyle{IEEEtran}
\bibliography{IEEE}

%


\begin{IEEEbiography}[{\includegraphics[width=1in,height=1.25in,clip,keepaspectratio]{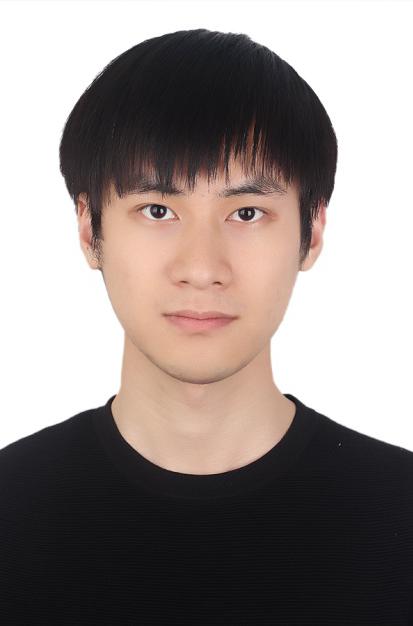}}]{Dihan Zheng}
is a Ph.D. student in in Yau Mathematical Sciences Center (YMSC), Tsinghua University supervised by Chenglong Bao. His research interests includes computer vision, image processing, generative modeling.
\end{IEEEbiography}

\begin{IEEEbiography}[{\includegraphics[width=1in,height=1.25in,clip,keepaspectratio]{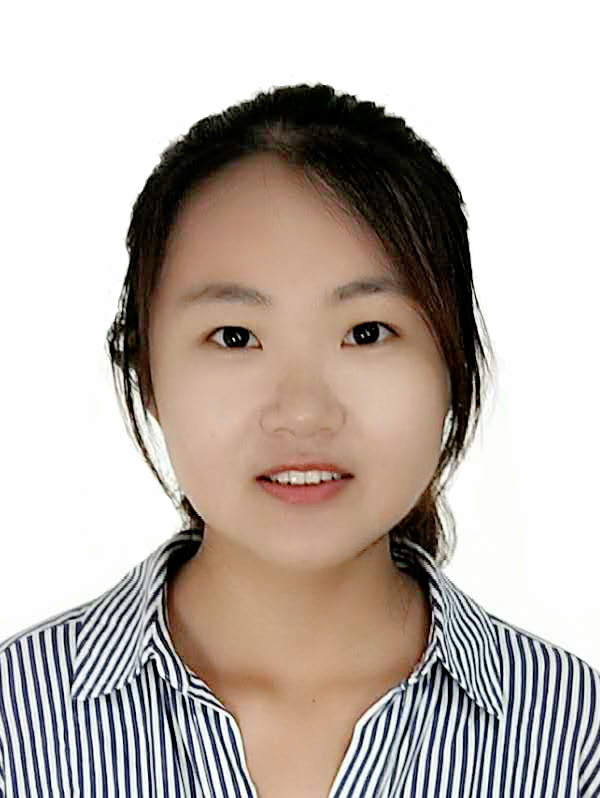}}]{Xiaowen Zhang}
is a senior engineer in hisilicon. She received her Ph.D. from information and communication engineering, Shanghai Jiao Tong University in 2019. Her main research interests include computer vision, image processing.
\end{IEEEbiography}

\begin{IEEEbiography}[{\includegraphics[width=1in,height=1.25in,clip,keepaspectratio]{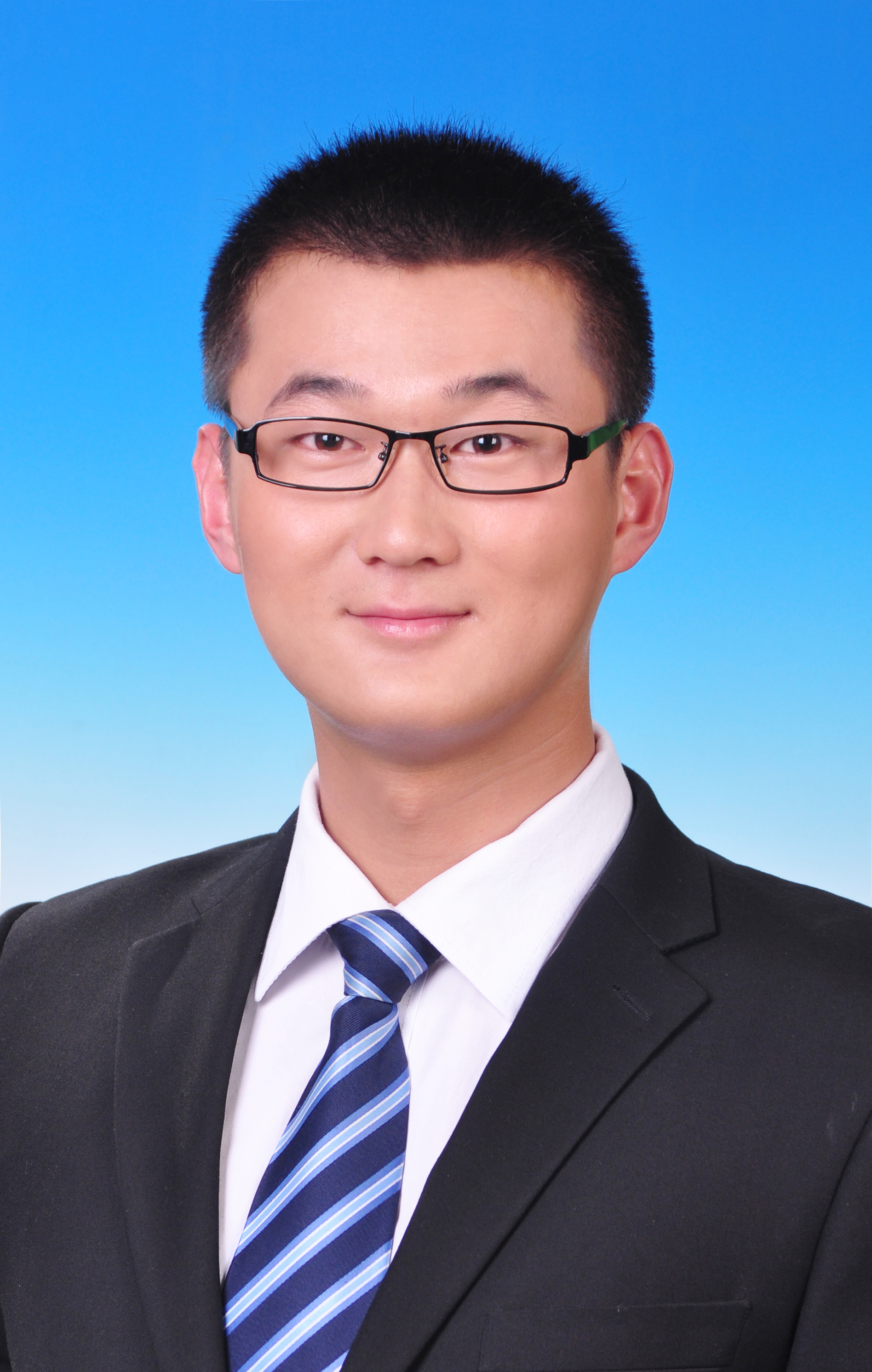}}]{Kaisheng Ma}
Kaisheng Ma is a tenure-track assistant professor of computer science in Institute for Interdisciplinary Information Sciences (IIIS), Tsinghua University. He leads the Algorithms, Architecture and Chipset Lab (ArChip). His research interests lie in the interdisciplinary fields of human-machine interface (HMI), brain-spired AI algorithm design, computer vision for self-driving, compact model design, computer architecture and microelectronics. He believes in vertical integration for systems, with a special emphasis on implanted medical and neural devices, and chip level system solutions for self-driving systems. He received his Ph.D. in Computer Science and Engineering at the Pennsylvania State University, following Professor Vijaykrishnan Nanrayanan @ Penn State, Yuan Xie @ UCSB, Jack Sampson @ Penn State. His previous research on Nonvolatile Processor and Energy Harvesting won 2018 EDAA Best Dissertation Award. He publishes papers on conferences including NeuralPS, ICCV, AAAI, CVPR, ISCA, ASPLOS, MICRO, HPCA, DAC etc. He has won many awards, including: 2015 HPCA Best Paper Award, 2016 IEEE MICRO Top Picks, 2017 ASP-DAC Best Paper Award. 2018 EDAA Best Dissertation Award. Dr. Ma has many honors, including 2016 Penn State CSE Department Best Graduate Research, 2016 Cover Feature of NSF ASSIST Engineering Research Center, 2011 Yang Fuqing \& Wang Yangyuan Academician Scholarship.
\end{IEEEbiography}


\begin{IEEEbiography}[{\includegraphics[width=1in,height=1.2in,clip,keepaspectratio]{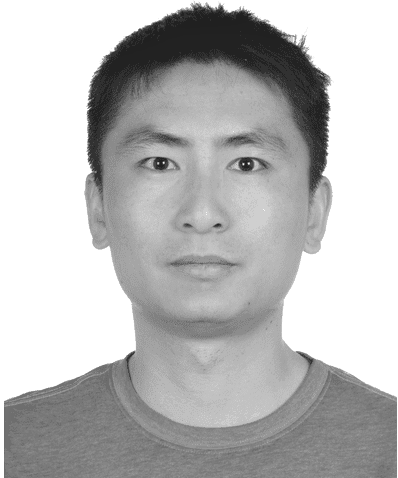}}]{Chenglong Bao} is an assistant professor in Yau mathematical sciences center, Tsinghua University and Yanqi Lake Beijing Institute of Mathematical Sciences and Applications. He received his Ph.D. from department of mathematics, National University of Singapore in 2014. His main research interests include mathematical image processing, large scale optimization and its applications.
\end{IEEEbiography}




\end{document}